\documentclass[a4paper,fleqn,10pt]{article}
\pdfoutput=1
\usepackage{amsmath}
\usepackage{amssymb}
\usepackage{array}
\usepackage{calc}
\usepackage{longtable}
\usepackage{multirow}
\usepackage{pstricks}
\usepackage{graphicx}
\usepackage{xspace}
\usepackage{cite}
\usepackage[a4paper,pdfborder={0 0 0}]{hyperref}
\usepackage[format=hang,labelfont=bf,hypcap=true]{caption}
\usepackage{subfig}
\usepackage{sectsty}
\allsectionsfont{\sffamily}
\subsubsectionfont{\mdseries\itshape\large}
\makeatletter
\DeclareRobustCommand*{\bfseries}{%
  \not@math@alphabet\bfseries\mathbf
  \fontseries\bfdefault\selectfont
  \boldmath
}
\makeatother
\let\spreprint\empty
\newcommand{\preprint}[1]{\def\spreprint{\protect#1}}
\let\sinstitute\empty
\newcommand{\institute}[1]{\def\sinstitute{\protect#1}}
\makeatletter
\renewcommand{\maketitle}{\begingroup
  \null\thispagestyle{empty}%
    \ifx\spreprint\empty
      \vskip 5ex
    \else
      \flushright\large\spreprint\vskip 2ex
    \fi
    \vskip 5ex
    \flushleft
      {\sffamily\bfseries
       \huge\@title
      }\vskip 6ex
      \@author\vskip 2ex
      \ifx\sinstitute\empty
      \else
        {\small\sinstitute}
      \fi
    \vskip 5ex
  \endgroup
}
\makeatother
\renewenvironment{abstract}{\begin{center}
  {\large\sffamily\bfseries Abstract: }
  \begin{minipage}[t]{0.75\textwidth}
}{\end{minipage}\end{center}\vskip 10ex}

\numberwithin{equation}{section}
\newcommand{\MCatNLO}{M\protect\scalebox{0.8}{C}@N\protect\scalebox{0.8}{LO}\xspace}

\newcommand{\POWHEG}{P\protect\scalebox{0.8}{OWHEG}\xspace}

\newcommand{\BlackHat}{B\protect\scalebox{0.8}{LACK}H\protect\scalebox{0.8}{AT}\xspace}

\newcommand{\Sherpa}{S\protect\scalebox{0.8}{HERPA}\xspace}

\newcommand{\CSS}{C\protect\scalebox{0.8}{SS}\xspace}

\newcommand{\LHC}{LHC\xspace}

\newcommand{\ATLAS}{ATLAS\xspace}
\newcommand{\CMS}{CMS\xspace}
\long\def\symbolfootnote[#1]#2{\begingroup%
\def\thefootnote{\fnsymbol{footnote}}\footnote[#1]{#2}\endgroup}
\newcommand{\EqRef}[1]{Eq.~\eqref{#1}}

\newcommand{\sbr}[1]{\left[ #1\right]}
\newcommand{\im}{\imath}
\newcommand{\jm}{\jmath}

\newcommand{\done}{{\rm d}}
\newcommand{\order}{\mathcal{O}}

\newcommand{\mr}[1]{\mathrm{#1}}

\newcommand{\nnb}{\nonumber}
\newcommand{\bea}{\begin{eqnarray}}
\newcommand{\eea}{\end{eqnarray}}
\newcommand{\bi}{\begin{itemize}}
\newcommand{\ei}{\end{itemize}}

\hypersetup{
  pdfauthor={Stefan Hoeche, Marek Schoenherr},
  pdftitle={Uncertainties in NLO + parton shower matched simulations of inclusive jet and dijet production}
}
\preprint{SLAC-PUB 15216\\IPPP/12/65\\DCPT/12/130\\LPN12-090\\MCNET-12-12}
\author{Stefan H{\"o}che$^1$, Marek Sch{\"o}nherr$^2$}
\title{Uncertainties in NLO + parton shower matched\\[3mm] simulations of inclusive jet and dijet production}
\institute{$^1$ SLAC National Accelerator Laboratory, 
  Menlo Park, CA 94025, USA\\
  $^2$ Institute for Particle Physics Phenomenology,
  Durham University, Durham DH1 3LE, UK\\}
\begin{document}
\maketitle
\begin{abstract}
We quantify uncertainties in the Monte-Carlo simulation of inclusive and
dijet final states, which arise from using the \MCatNLO technique for matching 
next-to-leading order parton level calculations and parton showers.
We analyse a large variety of data from early measurements at the LHC.
In regions of phase space where Sudakov logarithms dominate
over high-energy effects, we observe that the main uncertainty can be 
ascribed to the free parameters of the parton shower. In complementary
regions, the main uncertainty stems from the considerable freedom in the
simulation of underlying events.
\end{abstract}
%= text ===========================================
\section{Introduction}
\label{Sec:Introduction}

QCD jet production constitutes an important background in a variety of
searches for theories beyond the Standard Model~\cite{Chivukula:1991zk,
  Dixon:1993xd,Kilic:2008ub,Han:2010rf,
  Bai:2011mr,Schumann:2011ji}. 
At the same time, measurements of inclusive jet and dijet cross sections 
are used to constrain parton distributions~\cite{Guzzi:2011sv,Martin:2009iq,
  Ball:2012cx}
and to determine the value of the strong coupling~\cite{Affolder:2001hn,
  Abazov:2009nc,Wobisch:2011vm}. 
Despite the tremendous importance of QCD jet production, 
precise predictions of event rates and kinematics using higher-order 
perturbation theory remain challenging. Only up to four-jet final states 
have been computed at the next-to-leading order so far~\cite{Ellis:1990ek,
  Ellis:1992en,Giele:1993dj,Giele:1994gf,Kilgore:1996sq,
  Nagy:2001fj,Nagy:2003tz,Bern:2011ep}.
Phenomenologists therefore typically rely on the simulation
of high-multiplicity signatures by Monte-Carlo event generators.

The \ATLAS and \CMS experiments at the CERN Large Hadron Collider have 
recently measured inclusive jet and dijet production~\cite{Aad:2011tqa,
  Aad:2011fc,Aad:2011jz,Chatrchyan:2011wn,Khachatryan:2011zj,
  Khachatryan:2011dx,Chatrchyan:2011wm},
with many observables implicitly probing higher-order effects.
The outstanding quality of these data allows to validate and refine 
existing Monte-Carlo tools. The scope of this publication is a quantification
of related perturbative and non-perturbative uncertainties.

Calculating next-to-leading order QCD corrections to arbitrary
processes has become a highly automated procedure, limited only by
the capacity of contemporary computing resources. 
Infrared subtraction techniques~\cite{Frixione:1995ms,Catani:1996vz,Catani:2002hc} 
are implemented by several general-purpose matrix element 
generators~\cite{Gleisberg:2007md,Frederix:2008hu,Czakon:2009ss,
  Frederix:2009yq,Frederix:2010cj}. 
The computation of virtual corrections is tackled by a variety 
of dedicated programs~\cite{
  Denner:2005nn,Ossola:2006us,Ellis:2007br,Binoth:2008uq,Ossola:2008xq,Berger:2008sj,
  Ellis:2008ir,KeithEllis:2009bu,Berger:2010zx,Ita:2011wn,Denner:2010jp,
  Bredenstein:2010rs,Cullen:2011ac,Cullen:2011xs,Cascioli:2011va,Hirschi:2011pa}.
Turning the parton level result into a prediction at the particle level 
then requires a matching to the parton shower in order to implement resummation. 
Two methods have been devised to perform this matching procedure, 
the \MCatNLO~\cite{Frixione:2002ik} and the 
\POWHEG~\cite{Nason:2004rx,Frixione:2007vw} technique.
While both are formally correct at the next-to-leading order, they exhibit
subtle differences, which have been in the focus of interest 
recently~\cite{Hoeche:2011fd,Nason:2012pr}. 
We shall continue this study to some extent and perform a detailed comparison 
of scale uncertainties with resummation uncertainties as well as ambiguities arising 
from the Monte-Carlo simulation of the underlying event. We will employ the \MCatNLO
technique to match next-to-leading order parton-level results for dijet
production with the parton shower as implemented in the event generator 
\Sherpa~\cite{Gleisberg:2003xi,Gleisberg:2008ta}. Virtual corrections are obtained
from the \BlackHat library~\cite{Berger:2008sj,Berger:2009zg,Bern:2011ep}.
Earlier studies of inclusive jet and dijet production used the \POWHEG 
approach~\cite{Alioli:2010xa}. They exhibit a large dependence on the parton-shower 
and underlying-event model~\cite{Aad:2011tqa}.
We expect that the conclusions drawn from our study will also apply 
to the simulation provided in~\cite{Alioli:2010xa}, as the \MCatNLO 
and \POWHEG techniques are of the same formal accuracy.

The outline of this paper is as follows: Section~\ref{Sec:MCatNLO}
introduces the theoretical framework for our study, including a description
of the new developments in \Sherpa, which allow to perform the variation
of scales in a manner consistent with analytical resummation techniques.
Section~\ref{Sec:Results} presents results and discusses
the size and relative importance of the various types of uncertainties.
Section~\ref{Sec:Conclusions} contains some concluding remarks.

\section{The \MCatNLO matching method and its uncertainties}
\label{Sec:MCatNLO}

This section outlines the essence of the \MCatNLO technique for matching 
next-to-leading order matrix elements and parton showers. We follow the notation 
introduced in~\cite{Hoeche:2011fd} and report on an extension of the \MCatNLO implementation 
therein, which allows to vary resummation scales in a more meaningful way. 
We point out the free parameters of the \MCatNLO method, which will be used 
to obtain quantitative predictions in Sec.~\ref{Sec:Results}.

\subsection*{Notation}

In the following, $\mr{B}(\Phi_B)$ will be used to label Born squared matrix 
elements, defined on the Born phase space $\Phi_B$, which are summed/averaged 
over final-/initial-state spins and colours and include parton luminosities 
as well as symmetry and flux factors. Squared matrix elements of real emission 
corrections are denoted by ${\rm R}(\Phi_R)$. They are defined on the real-emission 
phase space $\Phi_R$. Virtual corrections, including collinear counterterms, 
are denoted by ${\rm\tilde{V}}$. Real and virtual corrections induce infrared
singularities of opposite sign, which cancel upon integration~\cite{Kinoshita:1962ur,Lee:1964is}.
In order to exploit this cancellation for the construction of 
Monte-Carlo event generators, subtraction formalisms are 
invoked~\cite{Frixione:1995ms,Catani:1996vz}, which introduce real subtraction 
terms ${\rm D}_{ij,k}^\text{(S)}$ and their corresponding integrated counterparts
${\rm I}_{\widetilde{\im\jm},\tilde{k}}^\text{(S)}$. 

In the \MCatNLO method, one defines additional Monte-Carlo counterterms,
${\rm D}_{ij,k}^\text{(A)}$, which represent the evolution kernels of the 
resummation procedure. These counterterms must necessarily have the correct 
infrared limit in order for the method to maintain full NLO accuracy~\cite{Hoeche:2011fd}.
Therefore, full colour and spin information needs to be retained. 
Ordinary parton-shower evolution kernels are recovered from the MC counterterms 
by taking the limit $N_c\to\infty$ and averaging over spins.

Away from the collinear limit, the form of MC counterterms is 
less constrained, which essentially presents a source of uncertainty of the 
\MCatNLO method. This particular uncertainty will not be addressed here 
as it can be reduced by matrix-element parton-shower merging at the next-to-leading 
order~\cite{Lavesson:2008ah,Gehrmann:2012yg,Hoeche:2012yf}. 

It is particularly useful to identify the MC counterterms with infrared subtraction terms, 
${\rm D}_{ij,k}^\text{(A)}={\rm D}_{ij,k}^\text{(S)}$, up to phase space
constraints. This procedure was advocated in~\cite{Hoeche:2011fd}, but the corresponding
implementation suffered from unknown integrals in the integrated subtraction terms 
in the case that the phase-space was parametrised in terms of parton-shower
evolution and splitting variable. In this publication, the problem is solved
by performing the integral of the remainder term 
${\rm D}_{ij,k}^\text{(A)}-{\rm D}_{ij,k}^\text{(S)}$ numerically.

\subsection*{The \MCatNLO method}

In terms of the above defined quantities, omitting flavour- and phase-space mappings,
the expectation value of an arbitrary infrared safe observable $O$ in the 
\MCatNLO method is given by~\cite{Frixione:2002ik} 
\begin{equation}\label{eq:mcatnlo}
\begin{split}
  \langle O \rangle
  \,=&\;
  \int\done\Phi_B\; \bar{\rm B}^\text{(A)}(\Phi_B)
  \Bigg[\,
	\Delta^\text{(A)}(t_c,\mu_Q^2)\,O(\Phi_B)
  \\&\hskip 31mm
	+\sum_{\{\widetilde{\im\jm},\tilde{k}\}}
	 \int\limits_{t_c}^{\mu_Q^2}\done\Phi_{R|B}^{ij,k}\;
	 \frac{{\rm D}^\text{(A)}_{ij,k}(\Phi_B,\Phi_{R|B}^{ij,k})}{{\rm B}(\Phi_B)}\,
	 \Delta^\text{(A)}(t,\mu_Q^2)\;O(\Phi_R)
  \,\Bigg]\\
  &\;{}
  +\int\done\Phi_R\;
   \left[{\rm R}(\Phi_R)
	 -\sum_{ij,k}{\rm D}_{ij,k}^\text{(A)}(\Phi_R)\,\Theta(\mu_Q^2-t)
   \right]\;
   O(\Phi_R)
   \vphantom{\frac{{\rm D}^\text{(A)}_{ij,k}(\Phi_B,\Phi_{R|B}^{ij,k})}{{\rm B}(\Phi_B)}}\;.
\end{split}
\end{equation}
Therein, Born phase space configurations $\Phi_B$ are assigned next-to-leading 
order weights according to
\begin{equation}\label{eq:def_bbar_a}
  \begin{split}
    \bar{\rm B}^{\rm(A)}(\Phi_B)
    \,=&\;
    {\rm B}(\Phi_B)+\tilde{\rm V}(\Phi_B)+
      \sum_{\{\widetilde{\im\jm},\tilde{k}\}}
          {\rm I}_{\widetilde{\im\jm},\tilde{k}}^{\rm(S)}(\Phi_B)\\
    &\qquad+\sum_{\{\widetilde{\im\jm},\tilde{k}\}}
      \int\done\Phi_{R|B}^{ij,k}\;
      \sbr{\,\mr{D}^\text{(A)}_{ij,k}(\Phi_B,\Phi_{R|B}^{ij,k})\,
	     \Theta(\mu_Q^2-t)
        -\mr{D}^\text{(S)}_{ij,k}(\Phi_B,\Phi_{R|B}^{ij,k})\,}\;.
  \end{split}
\end{equation}
The real-emission phase space associated with parton emission off 
an external leg $\widetilde{\im\jm}$ of the Born configuration $\Phi_B$
can be factorised as $\Phi_R=\Phi_B\cdot\Phi_{R|B}^{ij,k}$. In this context,
$k$ denotes spectator partons in the splitting $\widetilde{\im\jm},\tilde{k}\to i,j,k$, 
which are used to absorb the recoil when a splitting parton $\widetilde{\im\jm}$ is put 
on-shell in the subtraction procedure. The emission phase space, $\done\Phi_{R|B}^{ij,k}$, 
can be parametrised as $\done\Phi_{R|B}^{ij,k}\propto\done t\,\done z\,\done\phi$, 
i.e.\ in terms of an evolution variable $t$, a splitting variable $z$ and 
an azimuthal angle $\phi$. The evolution variable $t$ is usually identified with 
some transverse momentum, $\mr{k}_\perp^2$. In the above equations, we always 
assume $t=t(\Phi_{R|B}^{ij,k})=t(\Phi_R,\Phi_B)$.

We call \EqRef{eq:def_bbar_a} the next-to-leading order weighted Born cross section.
The sum over parton configurations and the integral over the emission phase space 
in \EqRef{eq:def_bbar_a} are both evaluated using Monte-Carlo methods, 
while keeping the Born phase space point, $\Phi_B$, fixed. 
Because of the choice ${\rm D}_{ij,k}^\text{(A)}={\rm D}_{ij,k}^\text{(S)}$,
the integrand varies only mildly. A single phase-space point is therefore sufficient 
to obtain a reliable estimate of the integral. Note that in~\cite{Hoeche:2011fd}
the corresponding integral was absent as the phase-space constraints on
${\rm D}_{ij,k}^\text{(A)}$ were chosen to be the same as on ${\rm D}_{ij,k}^\text{(S)}$.
While this method lead to fewer fluctuations of the MC integral, it severely restricted
the flexibility of the \MCatNLO and hampered the correct assessment of uncertainties.
In this publication we are able to lift this restriction while still maintaining
full next-to-leading order accuracy of the simulation through incorporating
full colour and spin information in the resummation.

The resummation procedure itself is encoded in the square bracket multiplying the 
NLO-weighted Born matrix element, $\bar{\rm B}^{\rm(A)}$,
on the first and second line of \EqRef{eq:mcatnlo}. Note that the square bracket 
is unitary by construction. The overall Sudakov factor $\bar{\Delta}^{\rm(A)}$ 
is defined as
\begin{equation}\label{eq:nbp_ps}
  \bar{\Delta}^{\rm(A)}(t,t')\,=\;\prod_{\{\widetilde{\im\jm},\tilde{k}\}}
  \bar{\Delta}^{\rm(A)}_{\widetilde{\im\jm},\tilde{k}}(t,t')\;,
\end{equation}
with the partial Sudakov factors given by
\begin{equation}\label{eq:nbp_ijk}
  \begin{split}
  \bar{\Delta}^{\rm(A)}_{\widetilde{\im\jm},\tilde{k}}(t,t')\,=&\;
  \exp\left\{-\sum_{i=q,g}
    \int\limits_t^{t'}\done\Phi_{R|B}^{ij,k}\;
    \frac{1}{S_{ij}}\,\frac{S_R}{S_B}\,
    \frac{\mr{D}^{\rm(A)}_{ij,k}(\Phi_B,\Phi_{R|B}^{ij,k})}
         {\mr{B}(\Phi_B)}\,\right\}\;.
  \end{split}
\end{equation}
Again, $t=t(\Phi_{R|B}^{ij,k})$ is implied and the dependence of the 
Sudakov factor on the Born phase space configuration is implicit.
This Sudakov factor differs from the ordinary parton-shower Sudakov by 
including full colour and spin correlations. Its implementation is 
detailed in~\cite{Hoeche:2011fd}. The factors $S_B$, $S_R$ and $S_{ij}$ account for 
the potentially different symmetry factors present in the Born and real-emission 
matrix elements and the parton-shower expression, respectively. In the latter, 
identical particles produced at different scales $t$ are distinguishable, 
leading to a factorisation of symmetry factors along the evolution chain.
The third line in \EqRef{eq:mcatnlo} encodes the non-logarithmic remainder
terms of the next-to-leading order real-emission correction. Subsequent parton-shower 
evolution is effected on both terms respecting the emission scales already present. 
If necessary, truncated parton-shower emission are inserted 
\cite{Nason:2004rx} to retain the logarithmic accuracy of the parton shower.

\subsection*{Uncertainties}

The evolution variable in the Monte-Carlo counterterms is limited from above 
by the resummation scale squared, $\mu_Q^2$. This scale was introduced 
in the context of analytic resummation~\cite{Dasgupta:2001eq,Dasgupta:2002dc,
  Banfi:2004nk,Bozzi:2005wk,Bozzi:2010xn}.
It can be used to assess the uncertainties associated with the resummation programme.
We will make extensive use of this possibility in Sec.~\ref{Sec:Results}.

When defining scales for different parts of the calculation, 
there are certain restrictions that have to be adhered to. In principle, both 
the factorisation scale, $\mu_F$, and the renormalisation scale, $\mu_R$, in
the NLO weighted Born ME and the hard remainder function can be chosen freely.
The difference induced by different scales is formally of $\order(\alpha_s^2)$ 
relative to the Born contribution. It can, however, have a sizable impact in practice.
 
A different scale can be chosen for the resummation kernel in the square bracket 
of \EqRef{eq:mcatnlo}, which corresponds to the scale employed by the parton shower. 
This scale is required to be consistent with the one used in the shower itself.
In order to achieve full next-to-logarithmic accuracy, it must be 
of the functional form of the relative transverse momentum of the splitting, 
$\mr{k}_\perp$, as $\mr{k}_\perp\to 0$ \cite{Amati:1980ch}.

Additional uncertainties arise from subsequent parton showers, hadronization
and the simulation of multiple parton scattering. We comment on these effects
in the following section.
\section{Results}
\label{Sec:Results}

In this section results generated with the \MCatNLO algorithm detailed previously 
are presented for inclusive jet and dijet production. Monte-Carlo predictions are 
compared to a wide variety of measurements made by both the \ATLAS and 
\CMS experiments at the \LHC at 7 TeV. The automated 
implementation of the \MCatNLO algorithm in the Monte-Carlo event generator \Sherpa, detailed 
in~\cite{Hoeche:2011fd}, is used where the only non-automated ingredient, the 
one-loop matrix element, is interfaced from \BlackHat 
\cite{Giele:1993dj,Giele:1994gf,Bern:2011ep} employing the methods of~\cite{Binoth:2010xt}.
Further QCD evolution is 
effected using \Sherpa's built-in \CSS parton shower \cite{Schumann:2007mg}. 
Non-perturbative corrections, including multiple parton interactions 
\cite{Alekhin:2005dx}, hadronization corrections 
\cite{Winter:2003tt,Krauss:2010xy} and hadron decays \cite{Krauss:2010xx}, 
are calculated using phenomenological models tuned to data. The standard 
tune for \Sherpa-1.4.0 has been used. Soft-photon corrections are simulated 
using \cite{Schonherr:2008av}. The CT10 parton distribution 
functions \cite{Lai:2010vv} have been employed throughout. All results are presented 
at the particle level, using only stable final state particles with a 
lifetime longer than 10~ps.

For the central theoretical prediction the scales have been chosen as 
\begin{center}
  $\mu_F=\mu_R=\tfrac{1}{4}\,H_T$
  \qquad\qquad\text{and}\qquad\qquad
  $\mu_Q=\tfrac{1}{2}\,p_\perp$\;.
\end{center}
Therein, $H_T$ is defined as the sum of the scalar transverse momenta 
of the jets found in the partonic process before applying any resummation. 
These partonic jets are defined using the anti-$k_\perp$ algorithm 
\cite{Cacciari:2008gp,Cacciari:2011ma} with $R=0.4$ and $p_\perp^\text{min}=20$ 
GeV. $p_\perp$ is the transverse momentum of any of the two jets of the Born 
phase space configuration upon which the \MCatNLO procedure is effected. 
To estimate the intrinsic uncertainty of the predictions the perturbative 
scales $\mu_F$ and $\mu_R$ have been varied independently by the 
conventional factor of 2 around the central scale while the 
resummation scale is kept fixed. For the resummation scale variation, 
taking into account the simple form of the exponent of the Sudakov factor, the prescription 
of \cite{Dasgupta:2002dc} is followed, varying $\mu_Q$ by a factor of 
$\sqrt{2}$ around the central choice while both $\mu_F$ and $\mu_R$ are kept 
fixed. Further, non-perturbative uncertainties, i.e.\ the impact of 
shortcomings of the phenomenological models, have been assessed following the 
prescription outlined in \cite{Maestre:2012vp}: Alternative tunes increasing 
and decreasing the mean charged multiplicity in the transverse region by 10\% 
were used to estimate the uncertainty in the multiple parton interaction 
model\footnote{
  This corresponds to changing the switch {\tt SIGMA\_ND\_FACTOR}$\mp$0.03 
  (increase/decrease) around the central tune value.
}. Exchanging the cluster hadronization model of 
\cite{Winter:2003tt,Krauss:2010xy} for the Lund string hadronization 
model \cite{Andersson:1983ia} has been found to have negligible impact 
on jet observables in previous studies \cite{Hoeche:2011fd,Maestre:2012vp}, 
and is therefore not considered here.

\begin{figure}[t]
  \centering
  \includegraphics[width=0.3\textwidth]{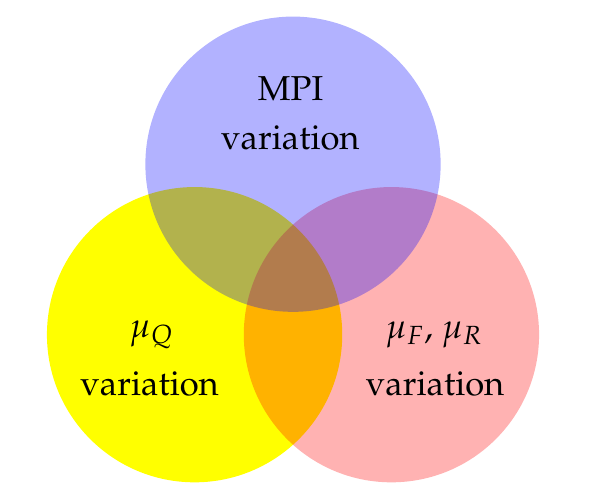}
  \caption{
           Colour scheme used to display various uncertainties. Overlapping 
           uncertainties will be displayed by adding the colours as indicated.
           \label{Fig:colour_scheme}
          }
\end{figure}

All observables studied in the following have been calculated using the 
same event sample. It is defined by requiring at least two anti-$k_\perp$ 
jets ($R=0.4$) with $p_\perp>10$ GeV at the parton level before applying 
any resummation, of which at least one must have $p_\perp>20$ GeV. Fig.\  
\ref{Fig:colour_scheme} presents the colour scheme used to display the 
individual uncertainties and their overlaps.

\begin{figure}[t]
  \centering
  \includegraphics[width=0.47\textwidth]{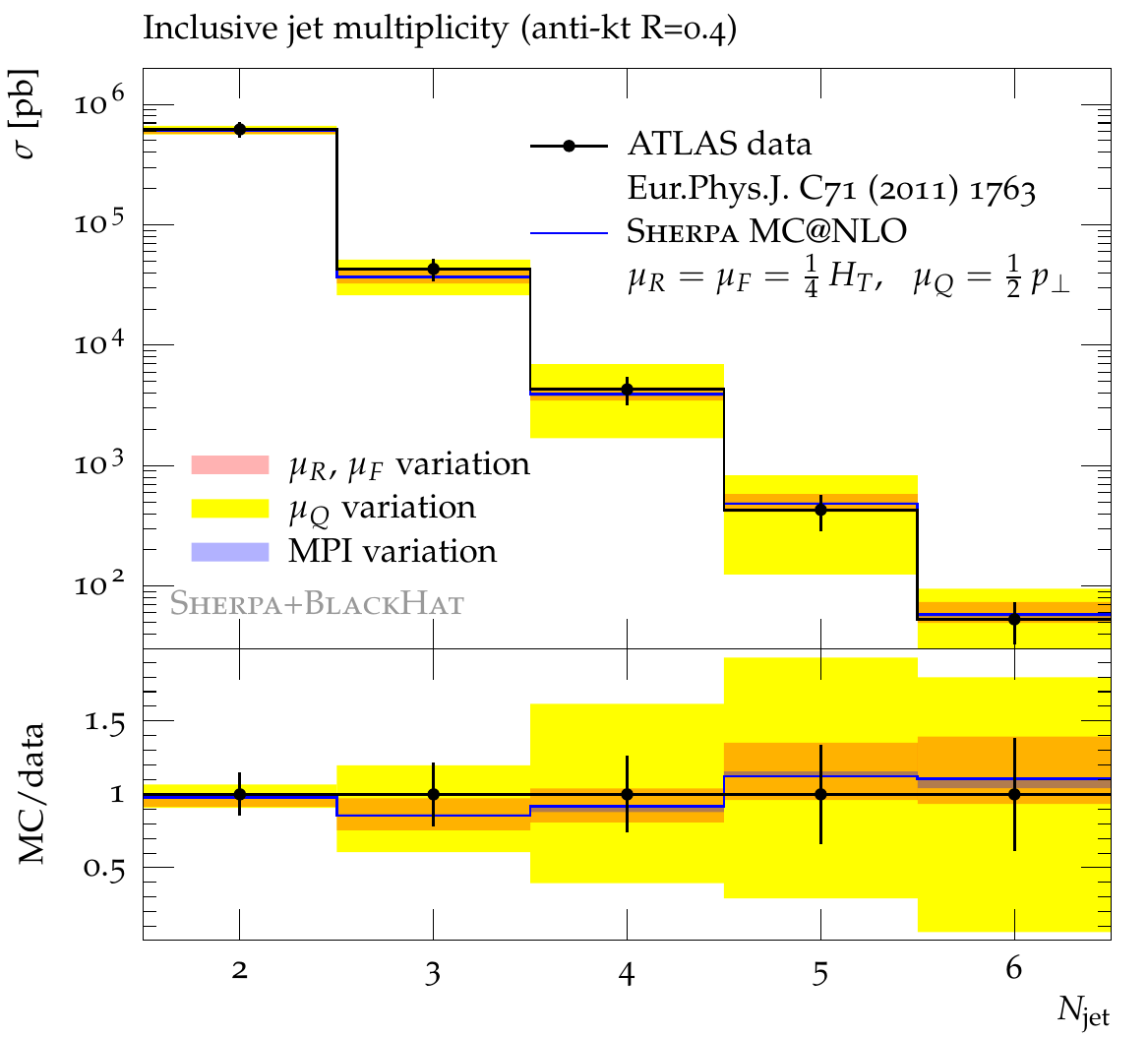}
  \vspace*{-2mm}
  \caption{
           Inclusive jet cross section compared to \ATLAS data \cite{Aad:2011tqa}.
           \label{Fig:ATLAS_inclusive_jet_xs}
          }
  \vspace*{-2mm}
\end{figure}

\subsection*{Inclusive jet rates} The first observables to study are inclusive 
jet production rates. These have been measured by the \ATLAS collaboration 
\cite{Aad:2011tqa}. Jets are defined at the particle level using 
the anti-$k_\perp$ algorithm with $R=0.4$ and $p_\perp>60$ GeV within 
$|y|<2.8$. Jets are ordered in transverse momentum.
Additionally, the leading jet is required to have $p_\perp>80$ GeV. 

Fig.\ \ref{Fig:ATLAS_inclusive_jet_xs} presents the results. We observe 
good agreement between our Monte-Carlo simulations and experimental data.
The renormalisation and factorisation scale uncertainty 
amounts to approximately 7\% for the dijet inclusive cross-section, which is 
described at next-to-leading order accuracy, while it increases to 14\% 
for the three-jet inclusive rate, which is described at leading order accuracy 
only. All higher multiplicity jet inclusive rates are described at the
logarithmic accuracy of the parton shower only and therefore inherit
the scale uncertainty of the inclusive three-jet rate. The resummation 
uncertainties, indicating the observables sensitivity to multiple higher-order
soft emissions below the resummation scale, is slightly larger: 8\% for 
the dijet inclusive cross section and 35\% for the three-jet inclusive rate. 
They steadily increase for higher jet multiplicities. The non-perturbative 
uncertainties, on the other hand, are negligible, contributing from $\sim$0.2\% 
for the dijet cross section to $\sim$6\% for the inclusive 5 jet cross section.

\begin{figure}[t]
  \vspace*{-2mm}
  \begin{minipage}{0.47\textwidth}
    \includegraphics[width=\textwidth]{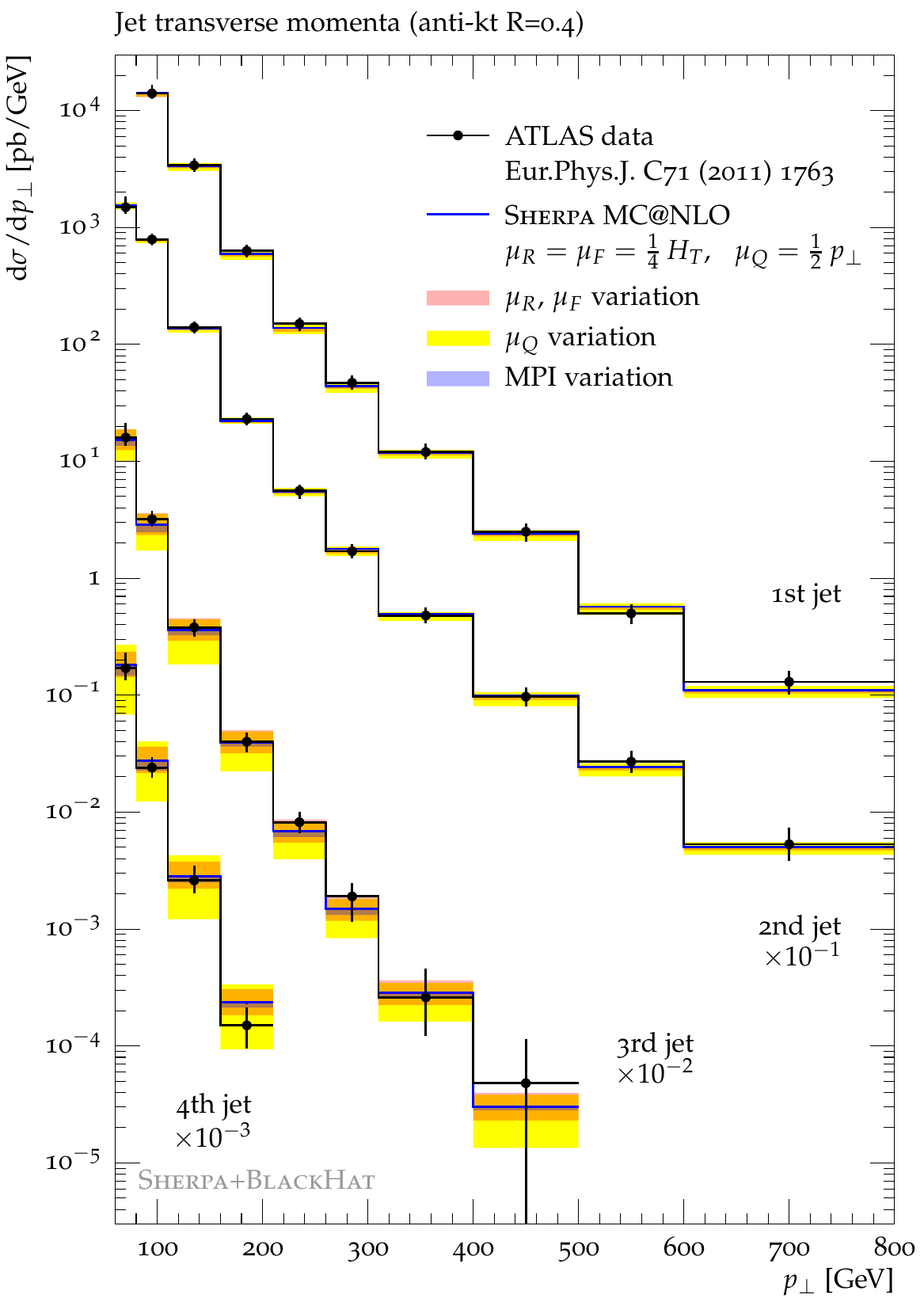}
  \end{minipage}
  \hfill
  \begin{minipage}{0.47\textwidth}
    \lineskip-1.85pt
    \includegraphics[width=\textwidth]{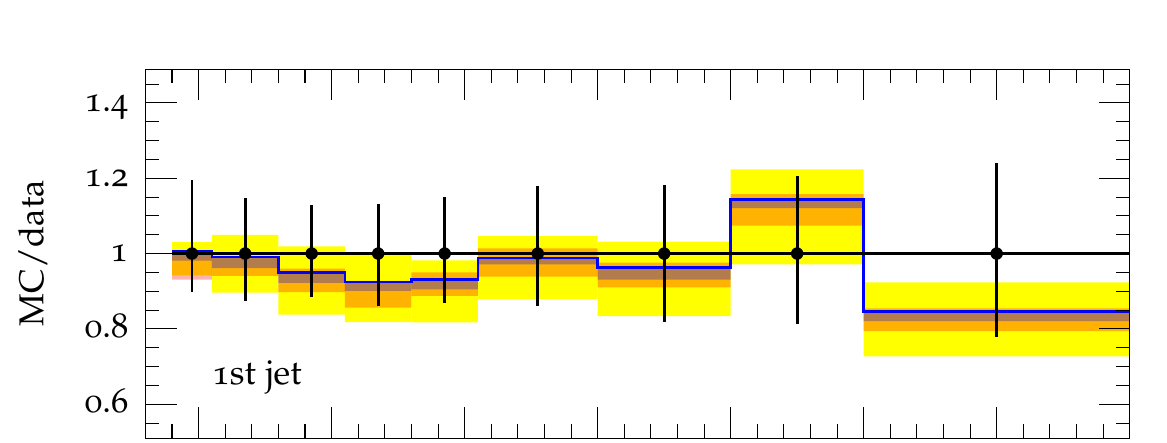}\\
    \includegraphics[width=\textwidth]{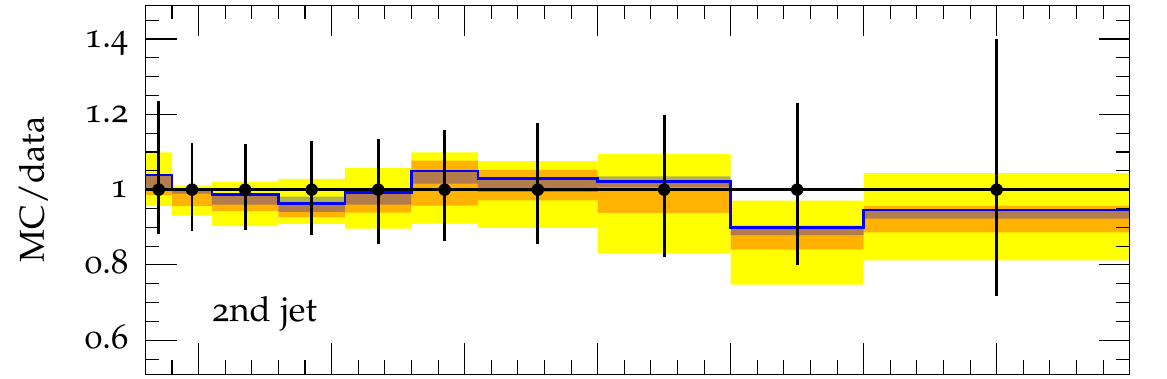}\\
    \includegraphics[width=\textwidth]{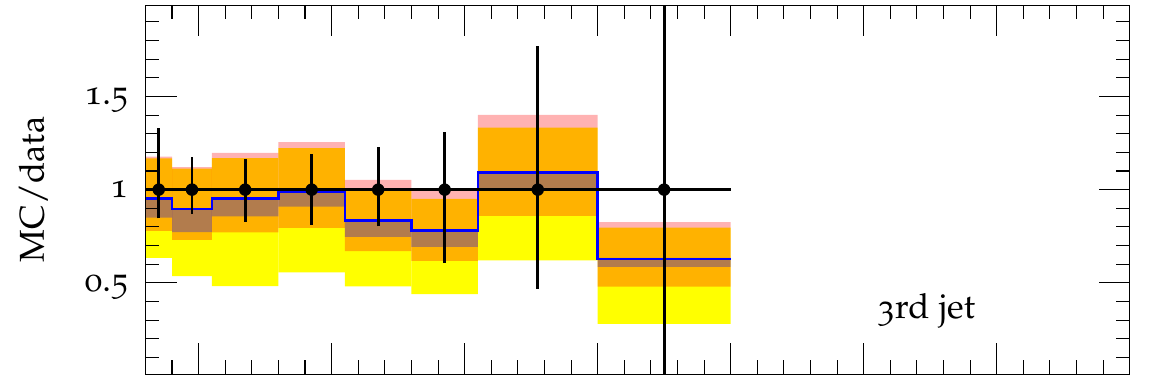}\\
    \includegraphics[width=\textwidth]{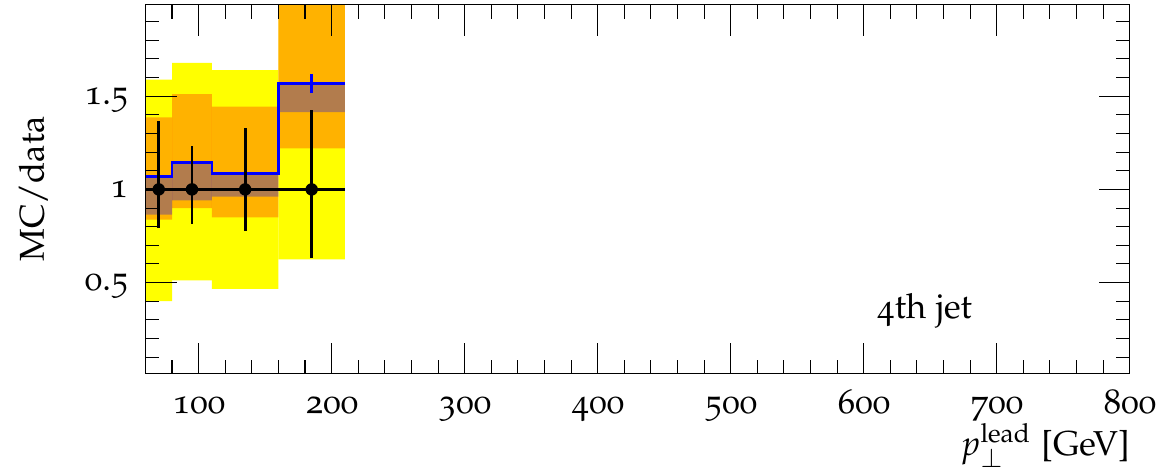}
  \end{minipage}
  \vspace*{-2mm}
  \caption{
           Jet transverse momenta compared to \ATLAS data \cite{Aad:2011tqa}.
           \label{Fig:ATLAS_jetpT}
          }
  \vspace*{-2mm}
\end{figure}

\begin{figure}[t!]
  \begin{minipage}{0.47\textwidth}
    \includegraphics[width=\textwidth]{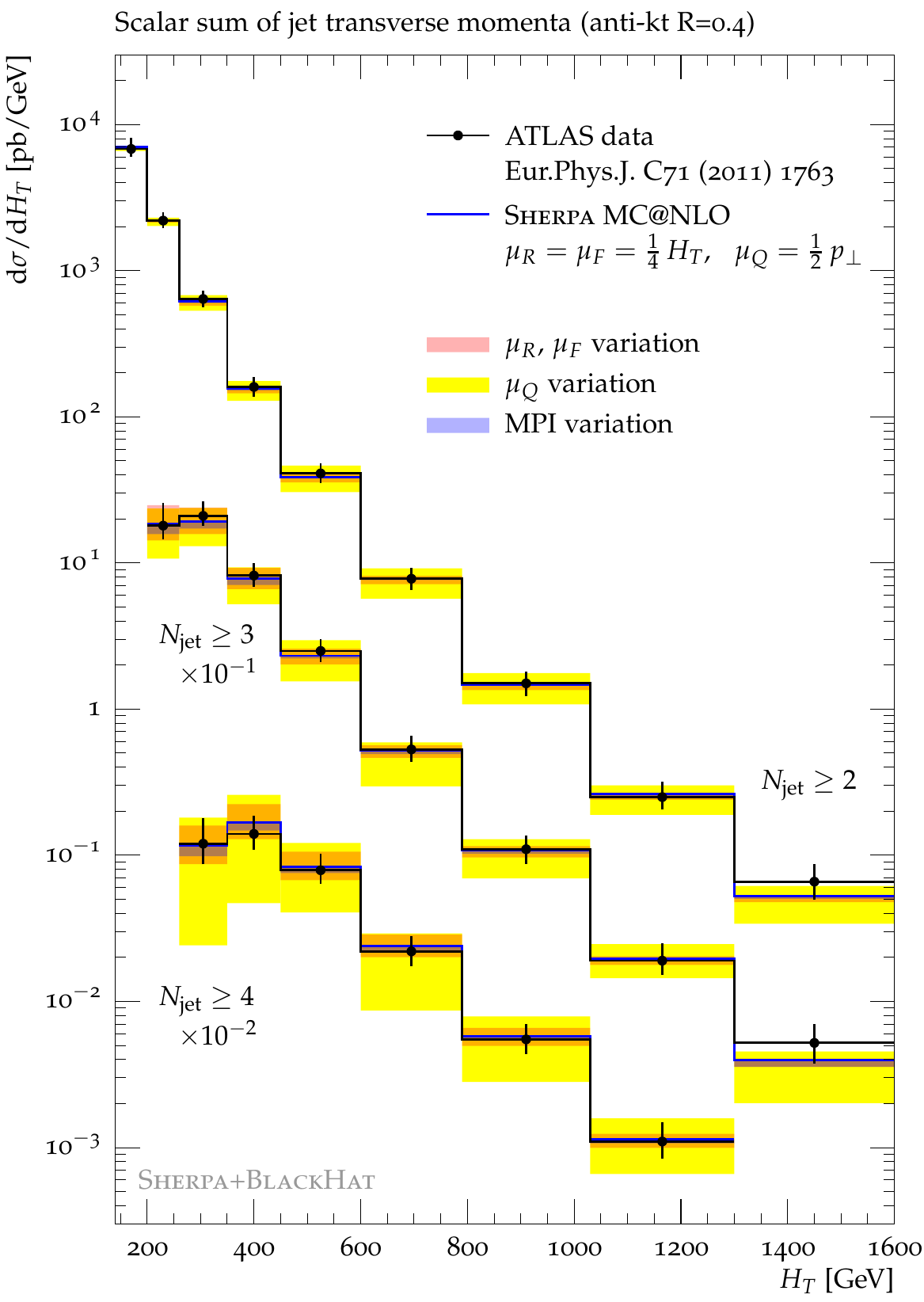}
  \end{minipage}
  \hfill
  \begin{minipage}{0.47\textwidth}
    \lineskip-1.85pt
    \includegraphics[width=\textwidth]{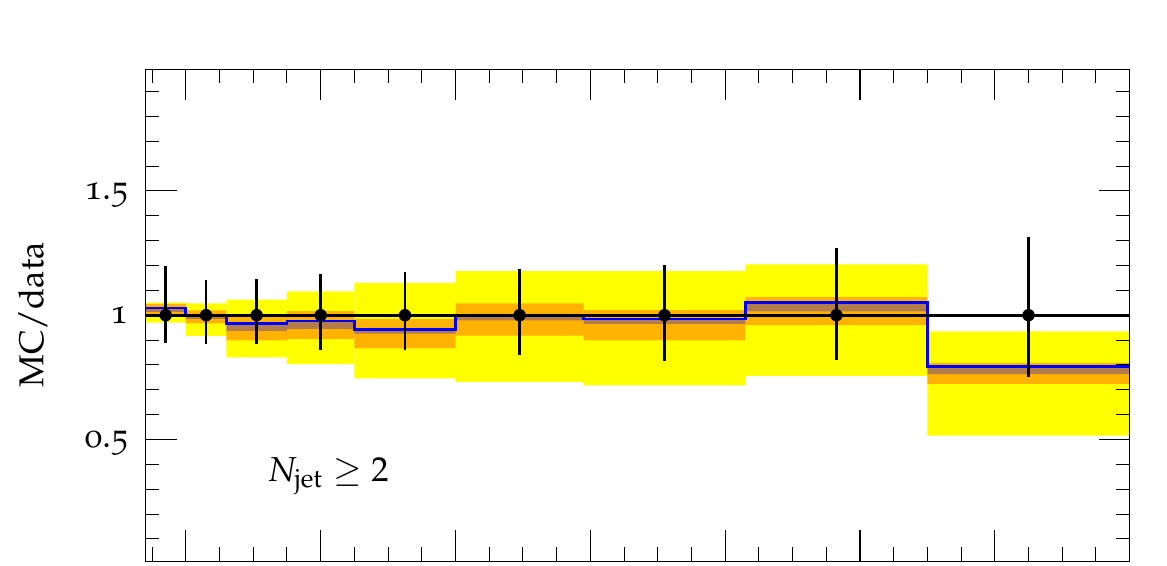}\\
    \includegraphics[width=\textwidth]{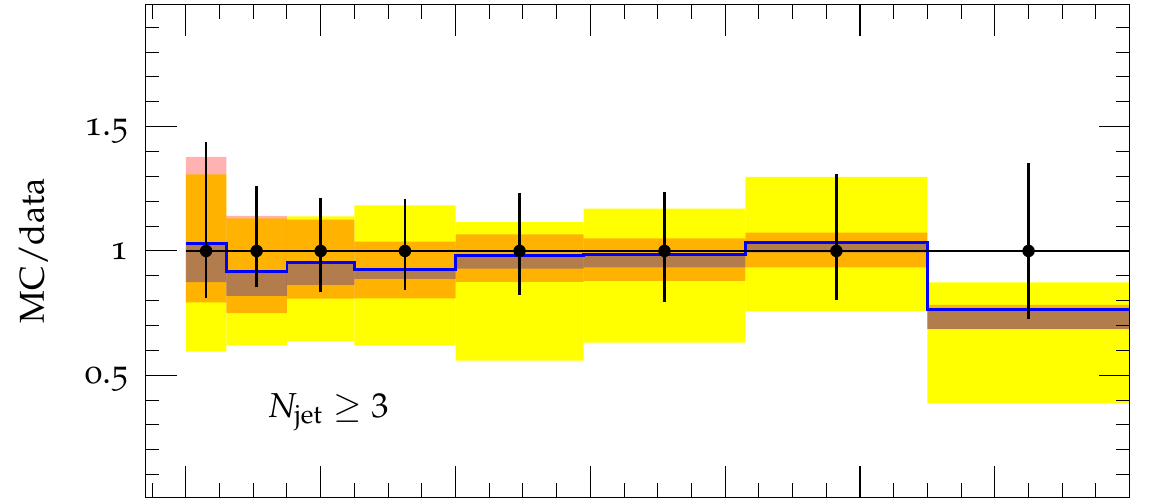}\\
    \includegraphics[width=\textwidth]{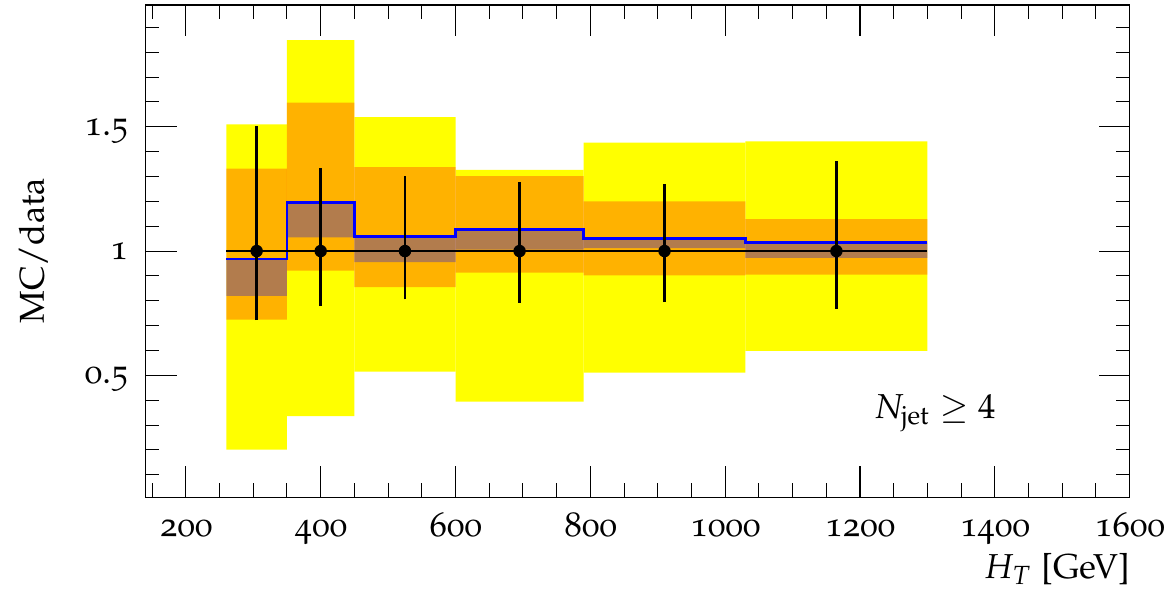}
  \end{minipage}
  \vspace*{-2mm}
  \caption{
           Scalar sum of jet transverse momenta compared to \ATLAS data 
           \cite{Aad:2011tqa}.
           \label{Fig:ATLAS_jetHT}
          }
  \vspace*{-2mm}
\end{figure}

\subsection*{Jet transverse momenta} 
The same analysis \cite{Aad:2011tqa} studied also the $p_\perp$-spectra of the 
individual jets. The event selection is the same as above, 
except that subleading jets with $p_\perp>60$ GeV may or may not be present
in case of the leading jet $p_\perp$. Fig.\ \ref{Fig:ATLAS_jetpT} 
displays the results compared to \ATLAS data. Again, we observe 
good agreement with our Monte-Carlo predictions. 
The two leading jets' transverse momenta, both calculated at next-to-leading 
order accuracy over large parts of the phase space, show the characteristically 
small renormalisation and factorisation scale dependencies of $\sim$5-10\%. 
Their resummation scale dependence is comparably very small throughout, 
ranging from $\sim$5\% at low $p_\perp$ to $\sim$20\% at large $p_\perp$.
This is expected from all choices of the resummation scale being smaller than 
the second jet's transverse momentum. Any 
influence therefore stems from the mismatch of (\MCatNLO) parton shower 
evolution and jet reconstruction. The transverse momentum of the third 
jet, being calculated at leading order accuracy, and the fourth jet, 
determined at leading logarithmic accuracy only, exhibit much larger 
scale variation, both for the renormalisation and factorisation scales and 
the resummation scale. Non-perturbative uncertainties are small in 
comparison. Similarly, Fig.\ \ref{Fig:ATLAS_jetHT} displays the 
scalar sum of the individual jet transverse momenta in events with at least two, three or 
four jets. For these observables good agreement is found as well. The 
perturbative and non-perturbative uncertainties are comparable to those of 
the individual jet transverse momenta.
\vspace*{-2mm}

\begin{figure}[t]
  \vspace*{-2mm}
  \includegraphics[width=0.47\textwidth]{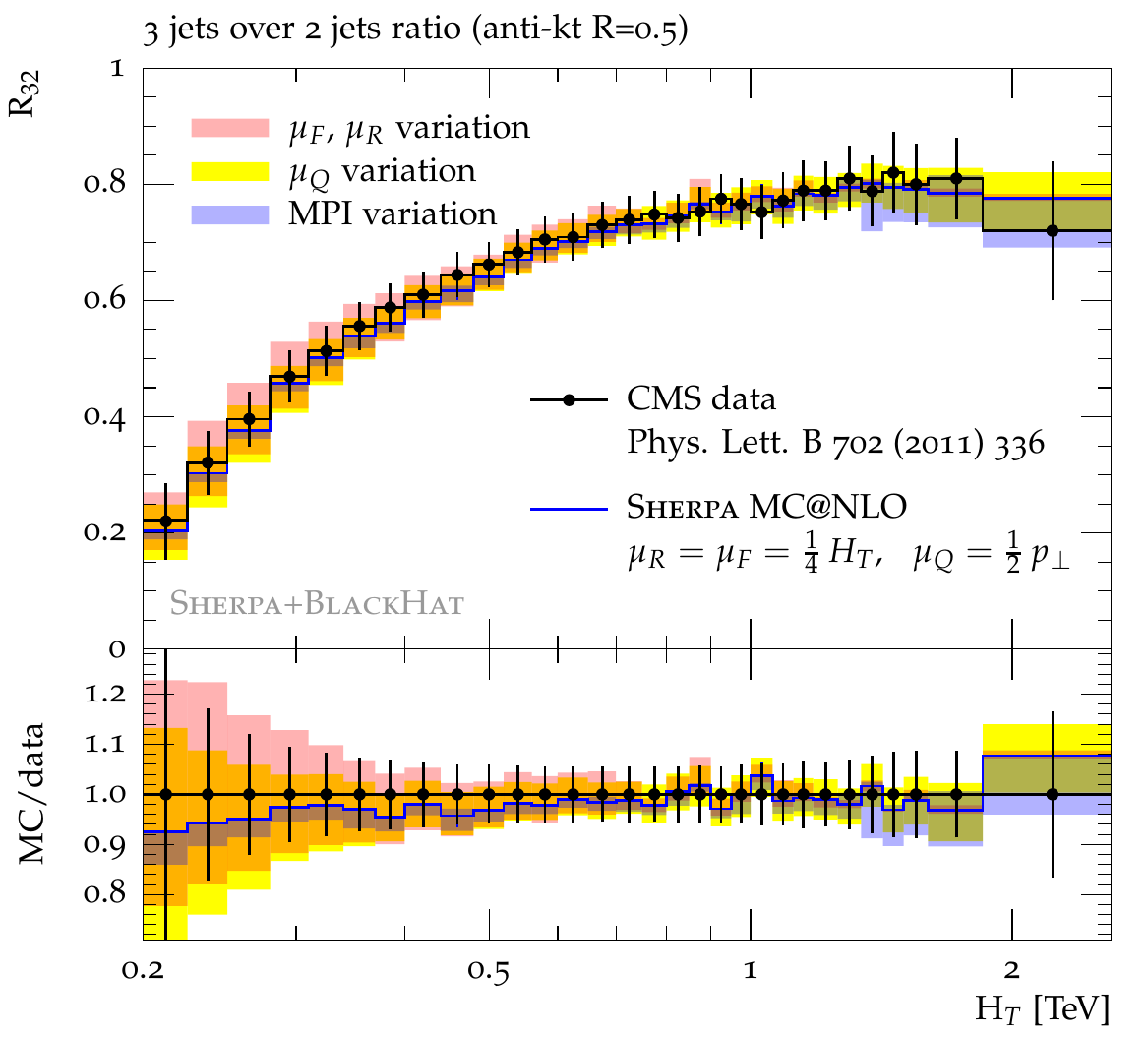}\hfill
  \includegraphics[width=0.47\textwidth]{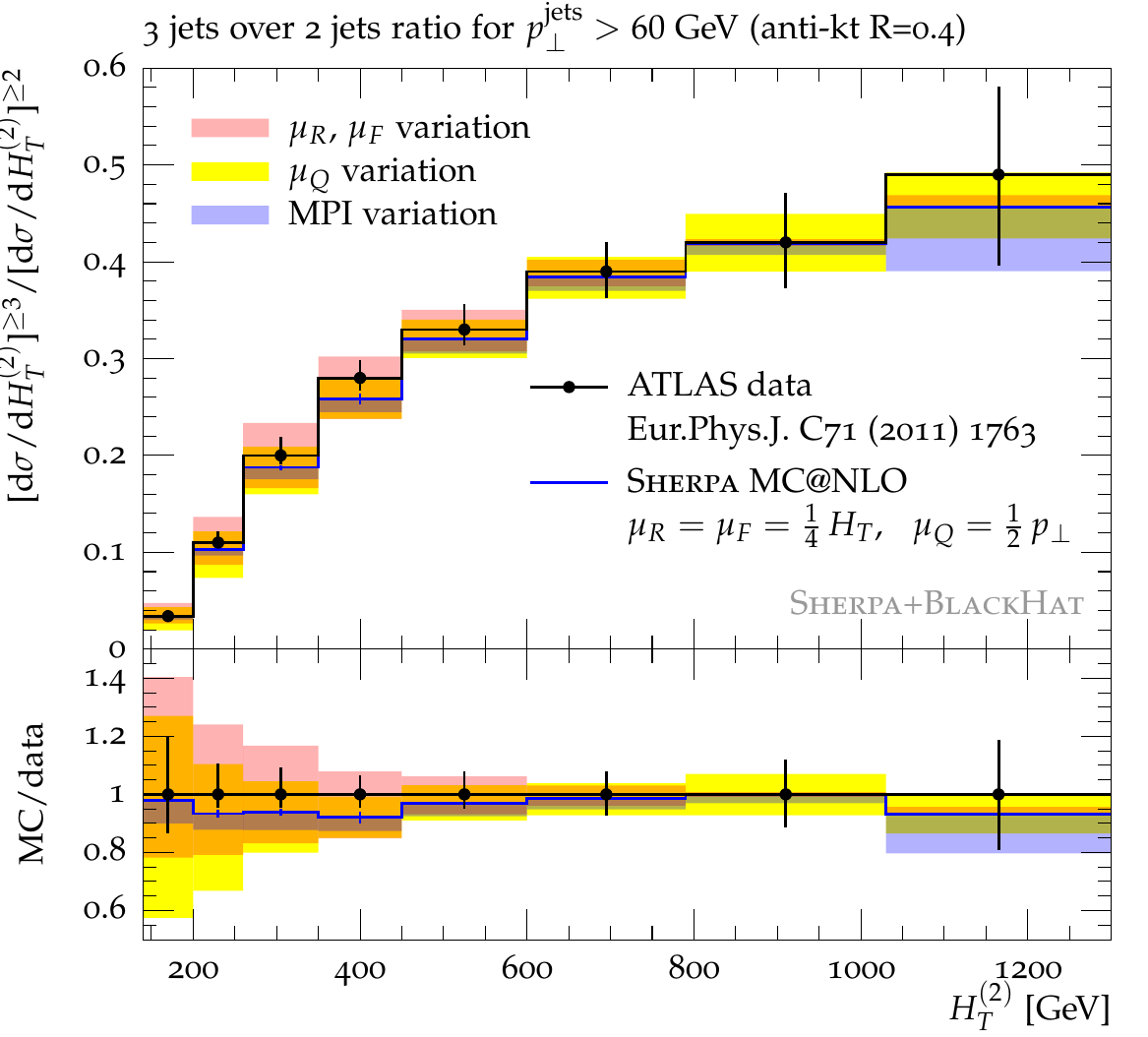}
  \vspace*{-2mm}
  \caption{
           3-jet over 2-jet ratio in dependence on the scalar transverse 
           momentum sum of all (\CMS) and the two leading (\ATLAS) jets 
           in comparison to \CMS \cite{Chatrchyan:2011wn} and \ATLAS data 
           \cite{Aad:2011tqa}.
           \label{Fig:CMS_ATLAS_R32_HT}
          }
  \vspace*{-2mm}
\end{figure}

\subsection*{3-jet over 2-jet ratio}
The next observable to be examined is the relative rate of inclusive three-jet 
events compared to inclusive two-jet events. The \CMS collaboration 
measured this ratio in dependence on the scalar sum of all jet transverse 
momenta, $H_T$, \cite{Chatrchyan:2011wn}. Within this 
analysis events with at least two and three jets, respectively, defined with the 
anti-$k_\perp$ algorithm with $R=0.5$, $p_\perp>50$ GeV, $|y|<2.5$, and
$H_T>0.2$ TeV were selected. The 3-jet over 2-jet ratio is then defined as 
$R_{32}=(\done\sigma_{\ge 3\text{jet}}/\done H_T)/
(\done\sigma_{\ge 2\text{jet}}/\done H_T)$. The comparison of the presented 
calculation to data is shown in the left panel of Fig.\ 
\ref{Fig:CMS_ATLAS_R32_HT}. Good agreement between MC predictions and data is observed. The scale 
variations in calculating $(\done\sigma_{\ge 3\text{jet}}/\done H_T)$, 
described at leading order, and $(\done\sigma_{\ge 2\text{jet}}/\done H_T)$, 
described at next-to-leading order, were done simultaneously because 
both observables were calculated from the same event sample. Consequently, the 
renormalisation and factorisation scale uncertainty are large at small 
$H_T$ but largely cancel at large $H_T$. The resummation uncertainty behaves 
similarly, but shows an opposite asymmetry at small $H_T$. The non-perturbative 
uncertainties are much smaller for small $H_T$ but grow to equivalent size 
in the large-$H_T$ region.

\begin{figure}[p]
  \vspace*{-5mm}
  \begin{minipage}{0.47\textwidth}
    \includegraphics[width=\textwidth]{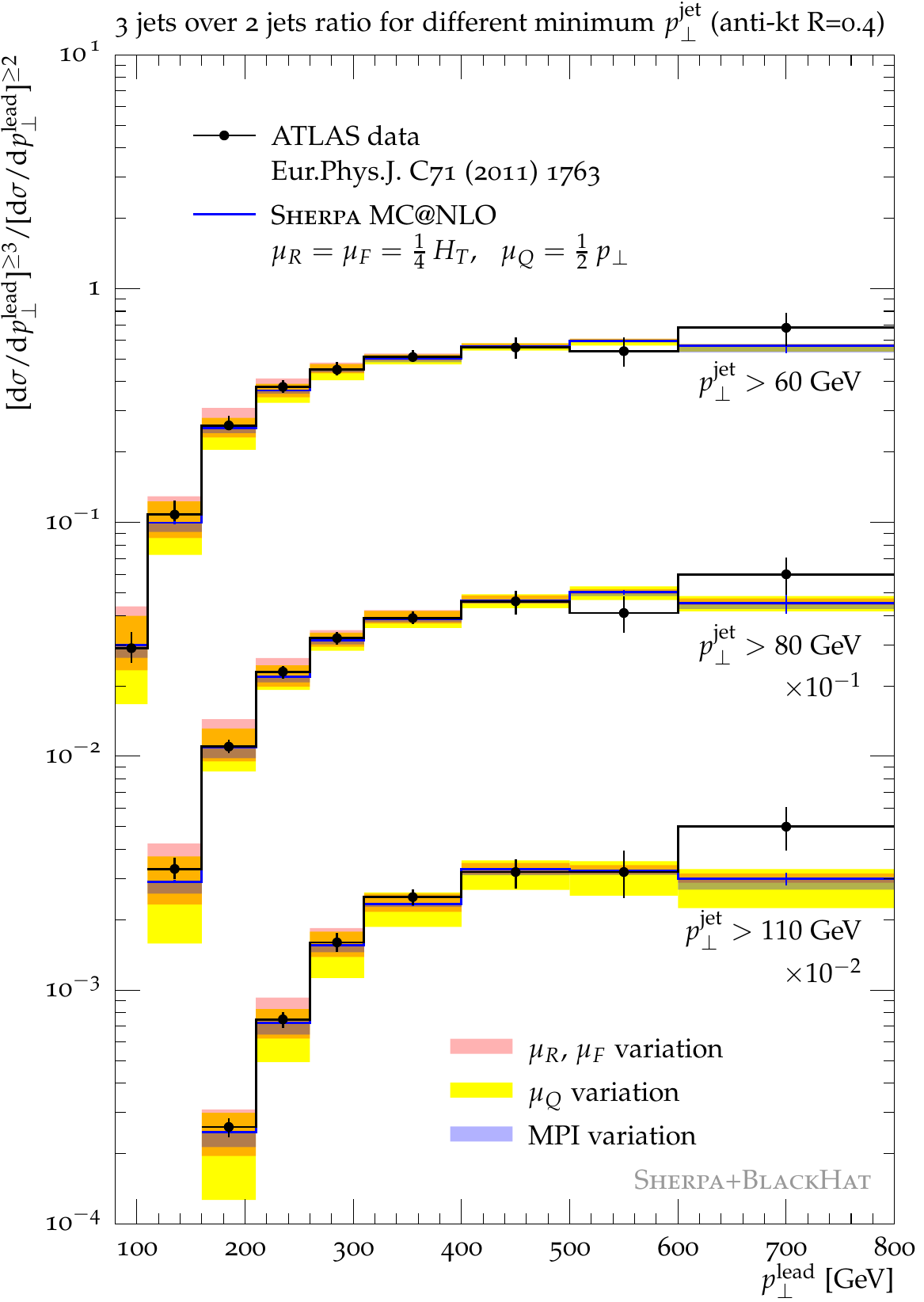}
  \end{minipage}
  \hfill
  \begin{minipage}{0.47\textwidth}
    \lineskip-1.85pt
    \includegraphics[width=\textwidth]{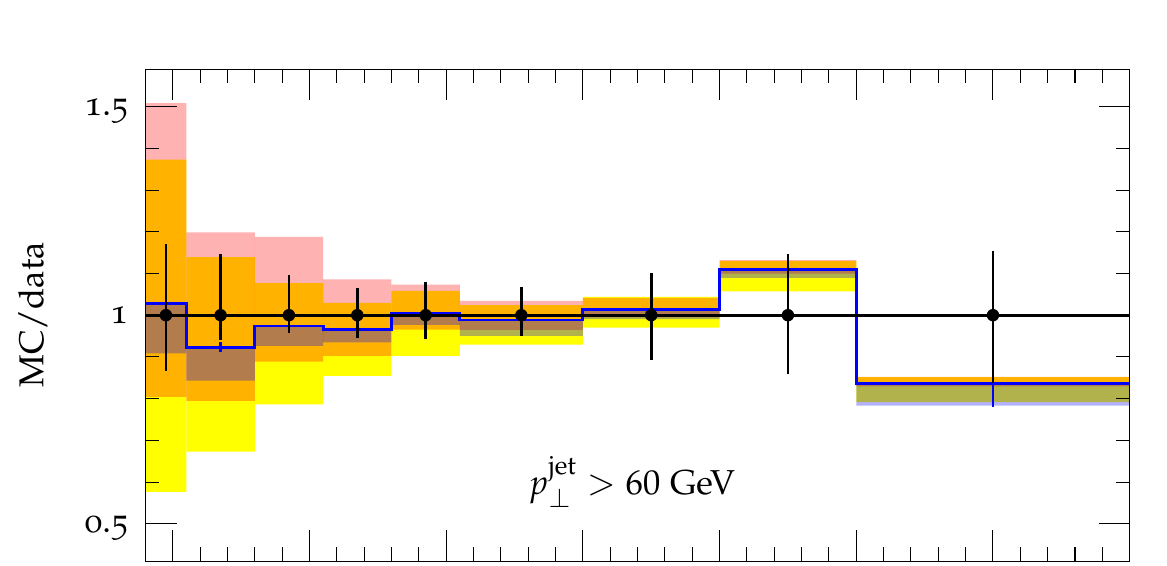}\\
    \includegraphics[width=\textwidth]{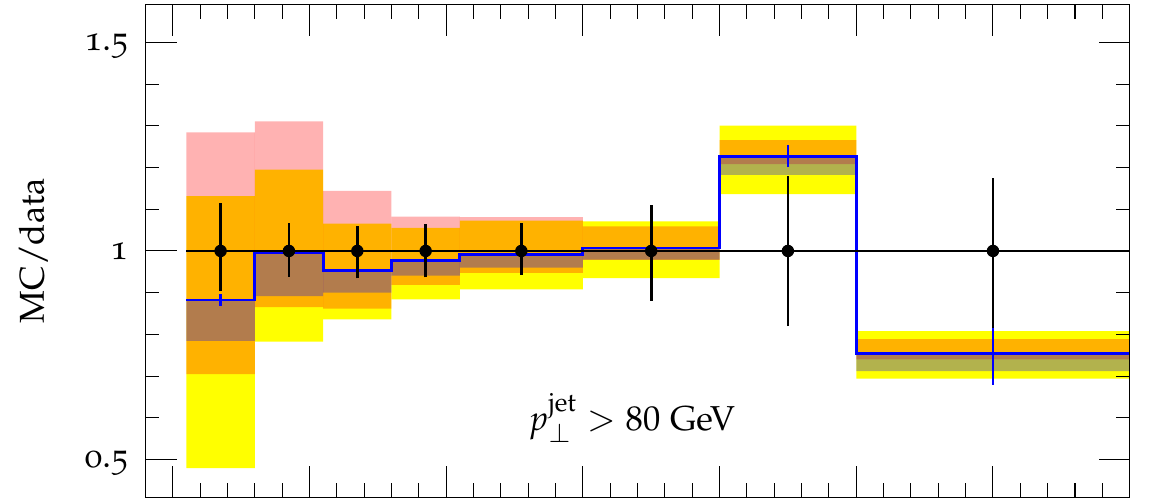}\\
    \includegraphics[width=\textwidth]{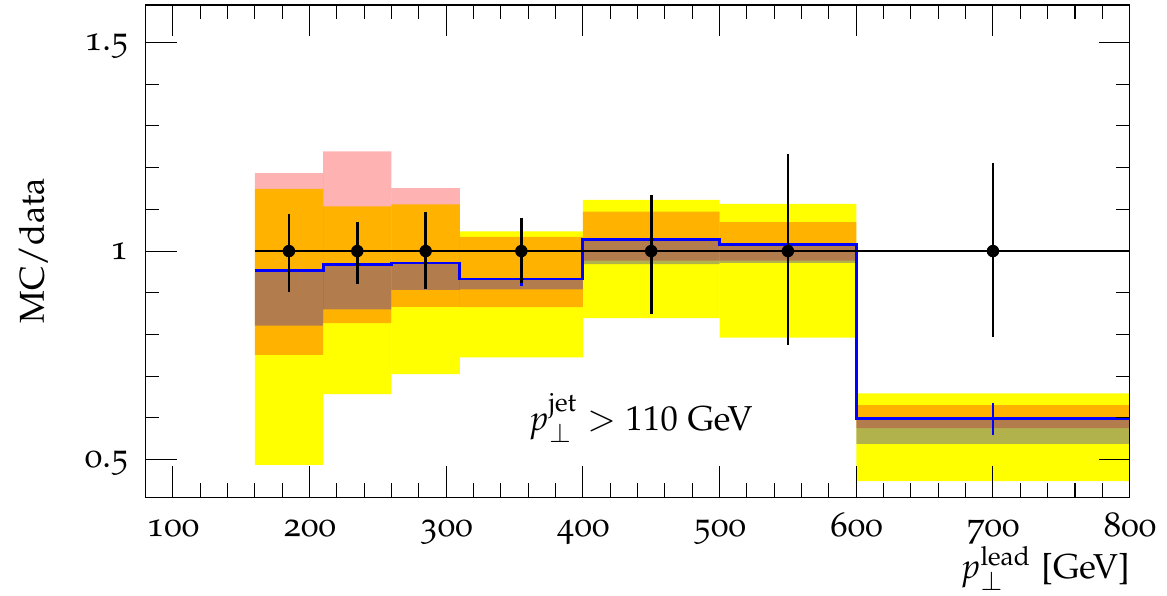}
  \end{minipage}
  \vspace*{-2mm}
  \caption{
           3-jet over 2-jet ratio in dependence on the leading jet transverse 
           momentum in comparison to \ATLAS data \cite{Aad:2011tqa}.
           \label{Fig:ATLAS_R32_pT}
          }
  \vspace*{-2mm}
\end{figure}

The analysis of \cite{Aad:2011tqa} also studied the 3-jet over 2-jet ratio 
both as a function of the the scalar transverse momentum sum of the two leading jets, 
$H_T^\text{(2)}$, and as a function of the transverse momentum of the leading jet, 
$p_\perp^\text{lead}$, only. The results are displayed in the right panel 
of Fig.~\ref{Fig:CMS_ATLAS_R32_HT} and in Fig.~\ref{Fig:ATLAS_R32_pT}, 
respectively. Both analyses show the same level of agreement between MC predictions and 
data. Scale uncertainties and non-perturbative uncertainties are of similar size as in
the \CMS analysis.

\begin{figure}[p]
  \begin{minipage}{0.47\textwidth}
    \includegraphics[width=\textwidth]{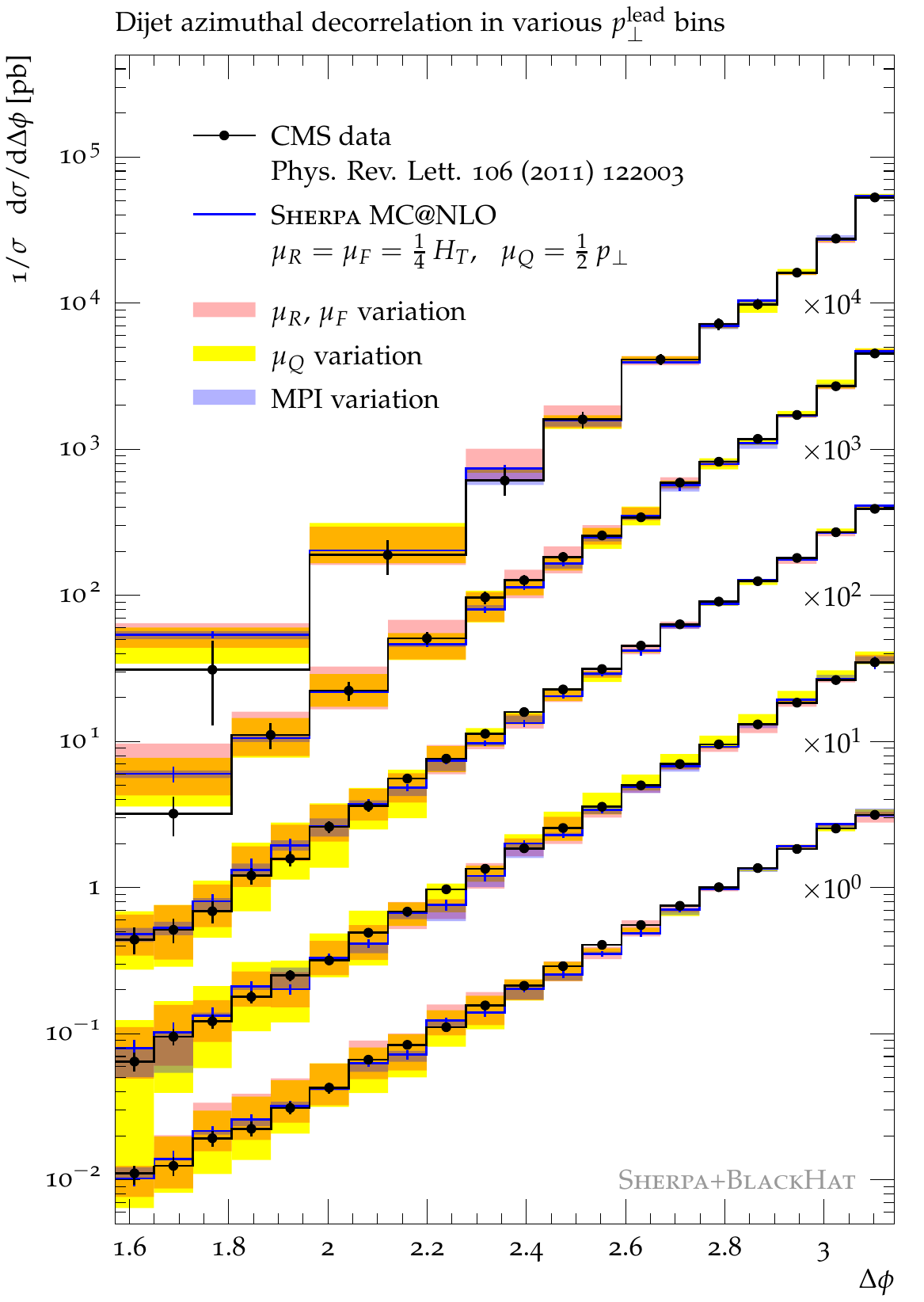}
  \end{minipage}
  \hfill
  \begin{minipage}{0.47\textwidth}
    \lineskip-1.85pt
    \includegraphics[width=\textwidth]{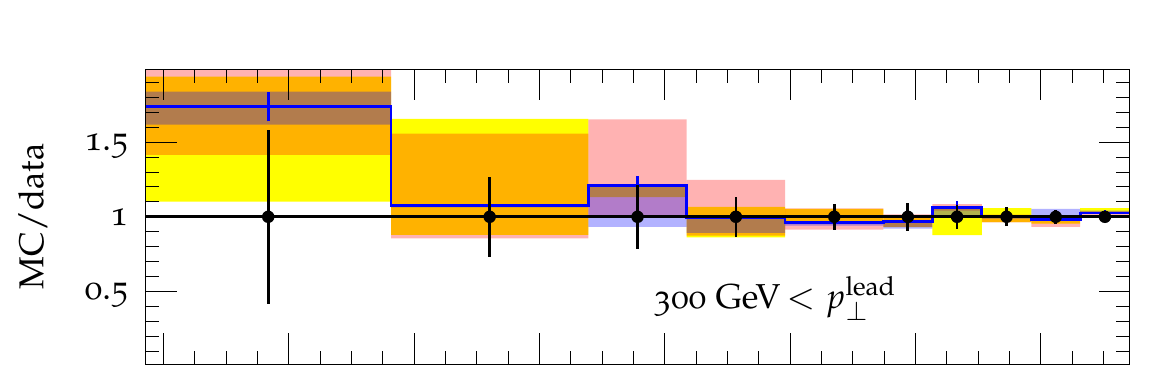}\\
    \includegraphics[width=\textwidth]{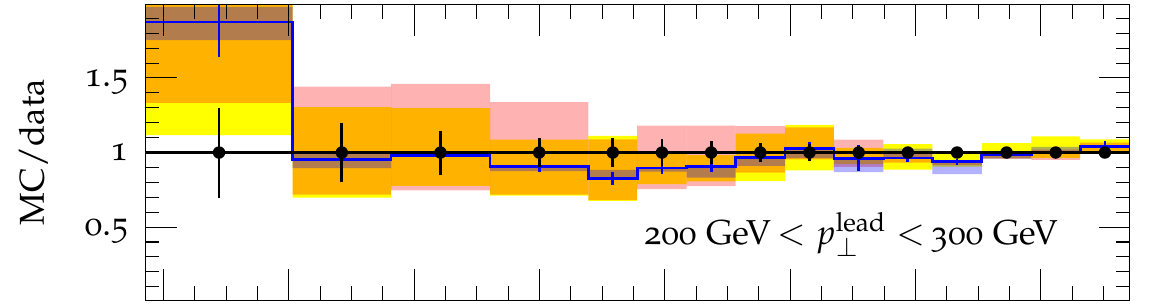}\\
    \includegraphics[width=\textwidth]{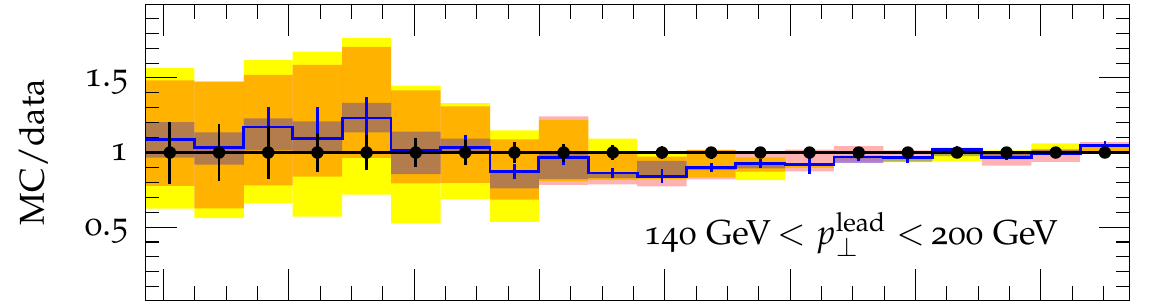}\\
    \includegraphics[width=\textwidth]{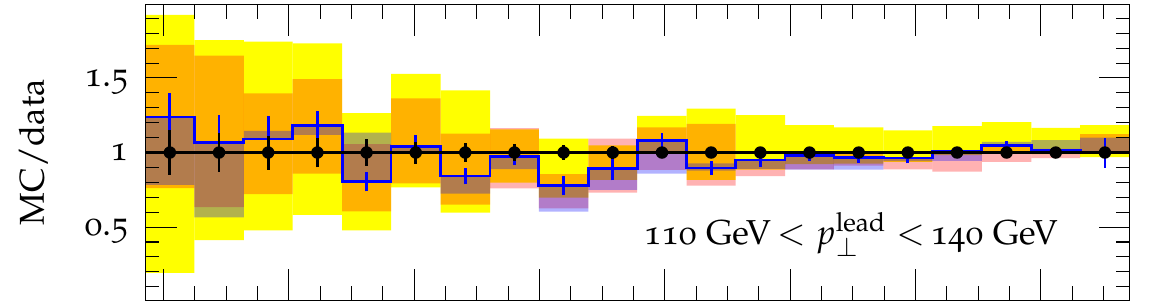}\\
    \includegraphics[width=\textwidth]{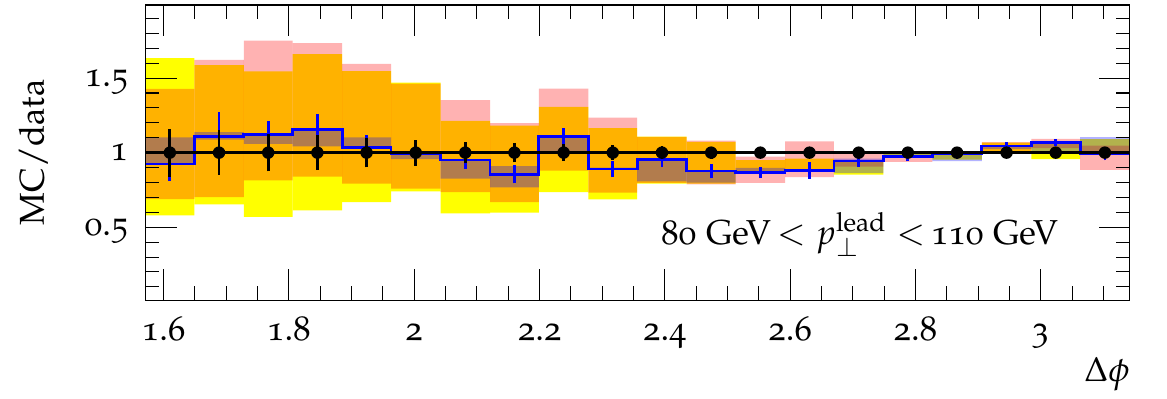}
  \end{minipage}
  \vspace*{-2mm}
  \caption{
           Azimuthal decorrelations of the leading jets compared to \CMS data 
           \cite{Khachatryan:2011zj}. The individual $p_\perp^\text{lead}$
           ranges on the left-hand side plot are ordered as indicated in the 
           right-hand side ratio plots. 
           \label{Fig:CMS_dijet_azimuthal_decorrelations}
          }
  \vspace*{-2mm}
\end{figure}

\subsection*{Azimuthal decorrelations}
Next, the correlations between the two leading jets are examined. The \CMS 
collaboration measured the dijet azimuthal decorrelations, i.e.\ the 
$\Delta\phi$ separation of the two leading jets, in \cite{Khachatryan:2011zj}. 
Therein, the jets are defined using the anti-$k_\perp$ jet algorithm with 
$R=0.5$ and a minimum transverse momentum of $p_\perp>30$~GeV within 
a rapidity interval of $|y|<1.1$. Events with at least two such jets were 
classified according to the leading jet's transverse momentum into five 
mutually exclusive regions: $p_\perp^\text{lead}\in[80,110]$ GeV, 
$[110,140]$ GeV, $[140,200]$ GeV, $[200,300]$ GeV, and $[300,\infty)$ GeV. 
The present calculation provides next-to-leading order accuracy at 
$\Delta\phi=\pi$, leading order accuracy in the region 
$\tfrac{2}{3}\pi<\Delta\phi<\pi$, and leading logarithmic accuracy in the 
region $\Delta\phi<\tfrac{2}{3}\pi$. The results are shown in 
Fig.~\ref{Fig:CMS_dijet_azimuthal_decorrelations}. Good agreement 
between data and MC prediction is found. 
The renormalisation and factorisation scale uncertainties are accordingly 
small in the region $\Delta\phi>\tfrac{2}{3}\pi$ and increase towards 
lower $\Delta\phi$ values. Varying the resummation scale leads to 
similarly large uncertainties, while non-perturbative uncertainties only 
play a minor role. However, by normalising the observable to the inclusive 
dijet cross section, the scale uncertainties are artificially reduced. 

\begin{figure}[t]
  \begin{minipage}{0.47\textwidth}
    \includegraphics[width=\textwidth]{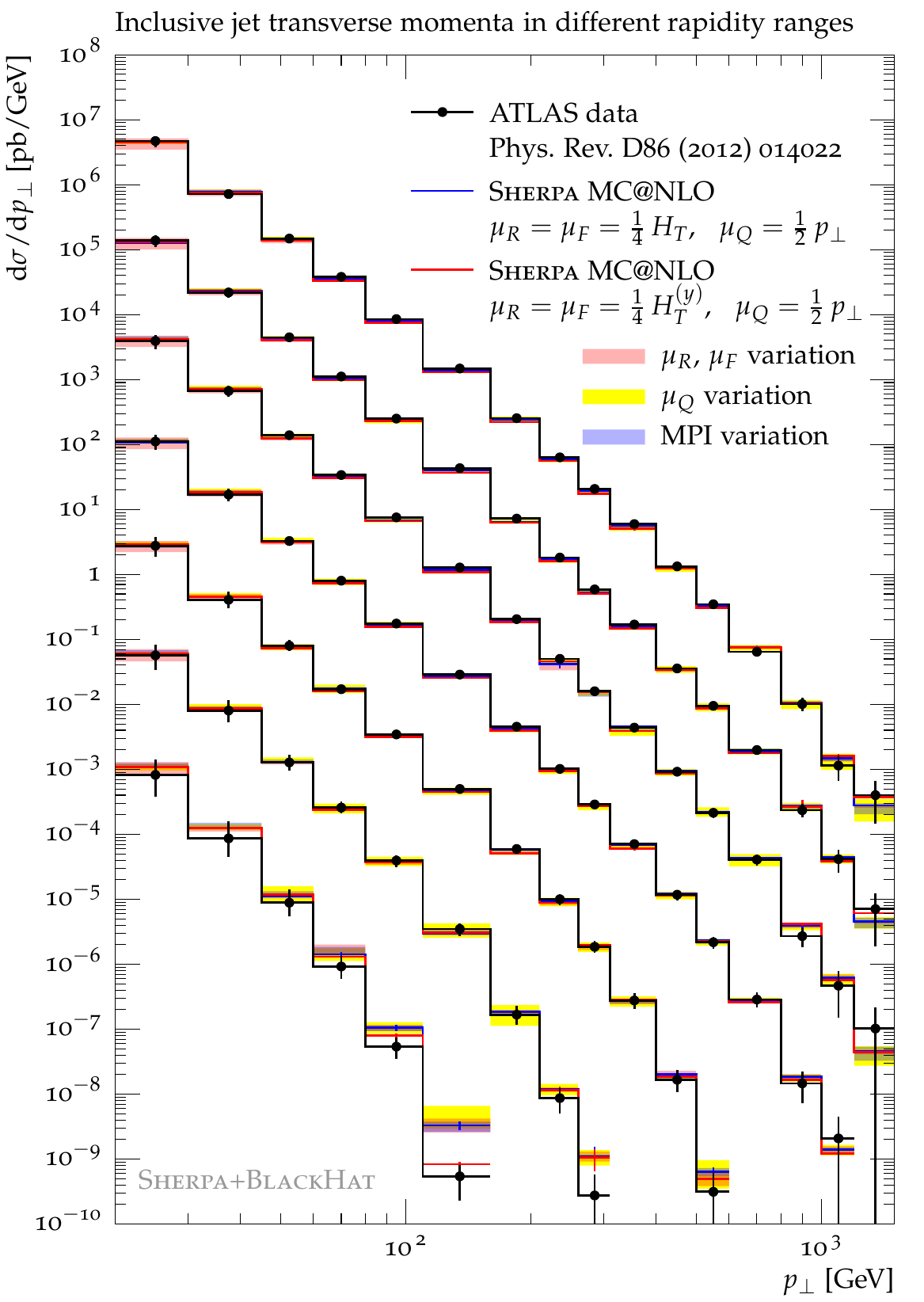}
  \end{minipage}
  \hfill
  \begin{minipage}{0.47\textwidth}
    \lineskip-1.85pt
    \includegraphics[width=\textwidth]{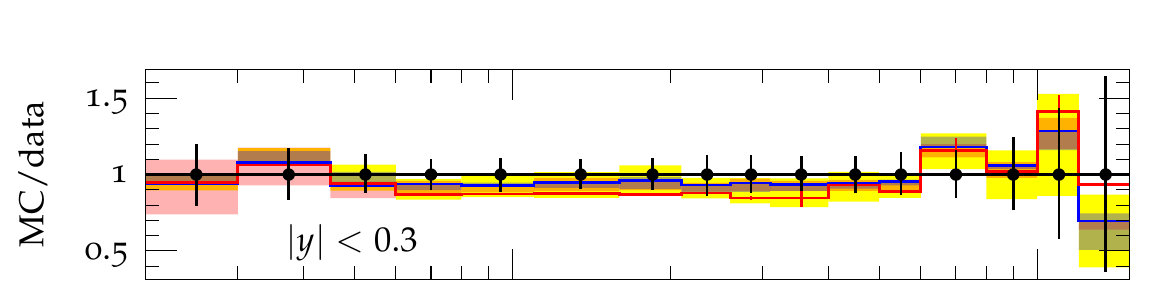}\\
    \includegraphics[width=\textwidth]{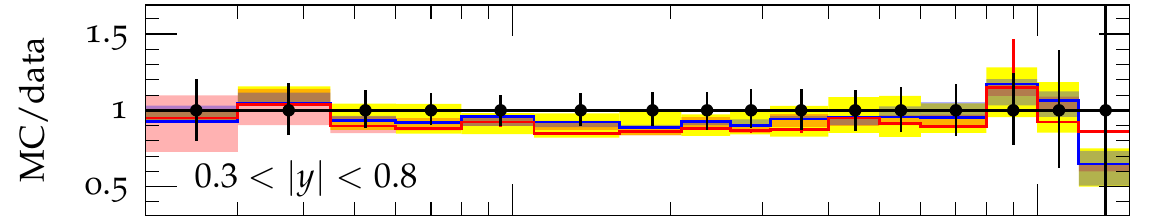}\\
    \includegraphics[width=\textwidth]{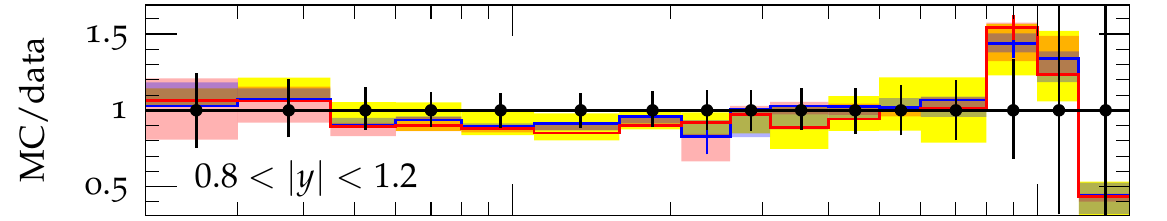}\\
    \includegraphics[width=\textwidth]{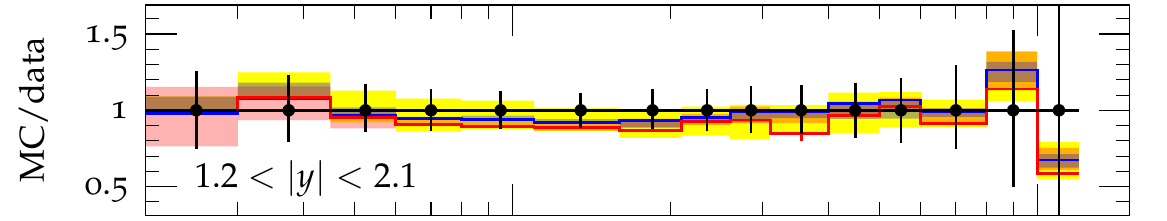}\\
    \includegraphics[width=\textwidth]{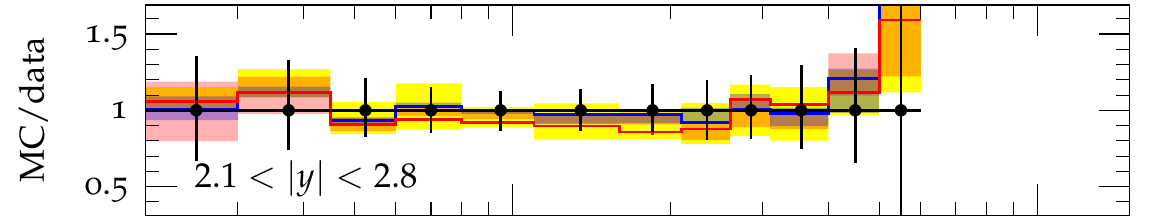}\\
    \includegraphics[width=\textwidth]{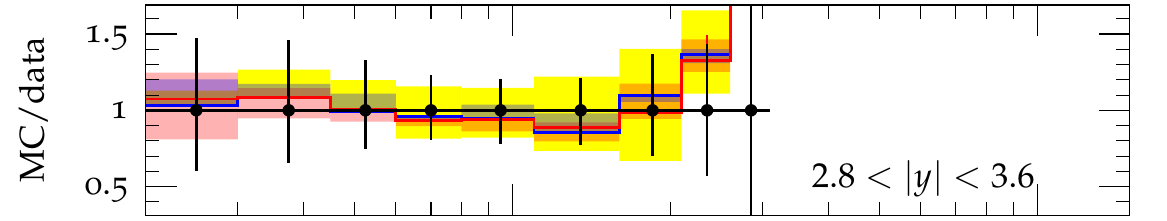}\\
    \includegraphics[width=\textwidth]{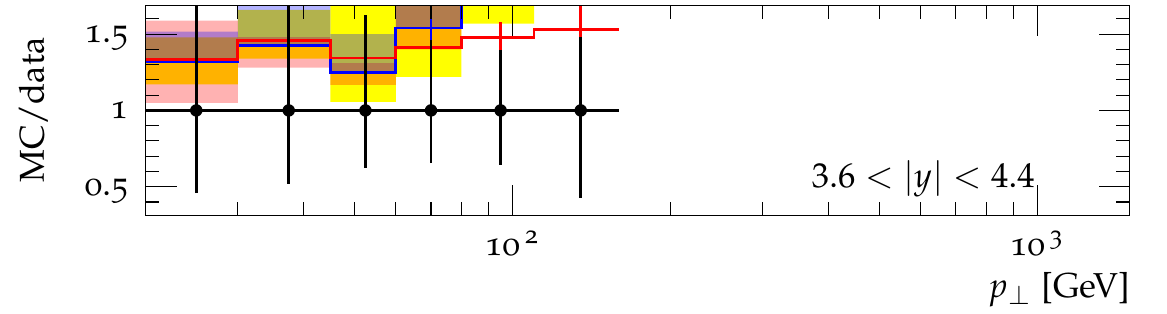}
  \end{minipage}
  \caption{
           Inclusive jet transverse momenta spectra different rapidity ranges 
           compared to \ATLAS data \cite{Aad:2011fc}. The different rapidity 
           ranges on the left-hand side plot are ordered as indicated in the 
           right-hand side ratio plots, being multiplied, from top to bottom 
           by factors of $1$, $3\!\cdot\!10^{-2}$, $10^{-3}$, 
           $3\!\cdot\!10^{-5}$, $10^{-6}$, $3\!\cdot\!10^{-8}$, and $10^{-9}$, 
           for readabilities sake. 
           \label{Fig:ATLAS_inclusive_jetpT}
          }
\end{figure}

\subsection*{Inclusive jet transverse momenta in different rapidity ranges}
To further study correlations between multiple jets produced, it is useful 
to consider double-differential observables studied by the \ATLAS 
collaboration \cite{Aad:2011fc}. We start with the inclusive jet transverse momentum in 
different rapidity ranges. Jets are defined using the anti-$k_\perp$ jet 
algorithm with $R=0.4$, $p_\perp>20$ GeV and $|y|<4.4$. 
Every jet is considered in the analysis. The contribution from the first two jets in the 
region where at least two jets are present is described at 
next-to-leading order accuracy, while the contribution of a possible third 
jet is described at leading order. All contributions of subsequent jets are 
described at leading logarithmic accuracy only. Thus, the overall accuracy 
of these observables is a mixture of the above.
Fig.~\ref{Fig:ATLAS_inclusive_jetpT} shows the result of the presented 
calculation compared to data. The agreement in all but the most forward 
rapidity ranges is good. Renormalisation and factorisation scale 
uncertainties are small for central jet production while they grow larger 
with increasing rapidity. The resummation scale uncertainty behaves similarly 
albeit being larger in magnitude throughout. Non-perturbative uncertainties 
are small, except in the very forward region, close to the beams. 
At very large rapidities the transverse momentum is no longer a good 
measure of the hardness of the process 
\cite{Ellis:1989vm,Ellis:1990ek,Ellis:1992en}. 
Instead a scale taking into account the dijet invariant mass should be used. 
Such consideration applied to an $H_T$-based scale, taking into account 
real emission dynamics, is proposed to take the following form 
\begin{equation}\label{eq:hty}
  \mu_{R/F}
  \;=\;\tfrac{1}{4}\,H_T^{(y)}
  \;=\;\tfrac{1}{4}\cdot\sum_{i\in\text{jets}} 
       |p_{\perp,i}|\,e^{f\,|y-y_\text{boost}|}
\end{equation}
wherein $y_\text{boost}=1/n_\text{jet}\cdot\sum_{i\in\text{jets}} y_i$, 
the rapidity of the $n$-jet system. The factor $f$ is chosen to suitably 
interpolate between the invariant-mass-like behaviour and 
transverse-momentum-like behaviour. For $f=0.3$ and the presence of only 
two jets it reduces exactly to the scale proposed in 
\cite{Ellis:1989vm,Ellis:1990ek,Ellis:1992en}, 
$\mu_{R/F}=\tfrac{1}{2}\,p_\perp\,e^{0.3 y^*}\approx m_{12}/(4\cosh(0.7 y^*))$. 
This scale choice, however, is only beneficial for describing data in this 
and the following analysis and either shows no impact or even reduces the 
agreement with the observed experimental data in all other analyses 
considered in this publication\footnote{
  Taking $\mu_{R/F}=\tfrac{1}{4}\,H_T^{(y)}$ as the central scale leads 
  to an underestimation of the inclusive three-jet rate presented in Fig.\ 
  \ref{Fig:ATLAS_inclusive_jet_xs} by 30\%, for example.}. 
It is therefore not adopted as a central scale choice.

\begin{figure}[t!]
  \begin{minipage}{0.47\textwidth}
    \includegraphics[width=\textwidth]{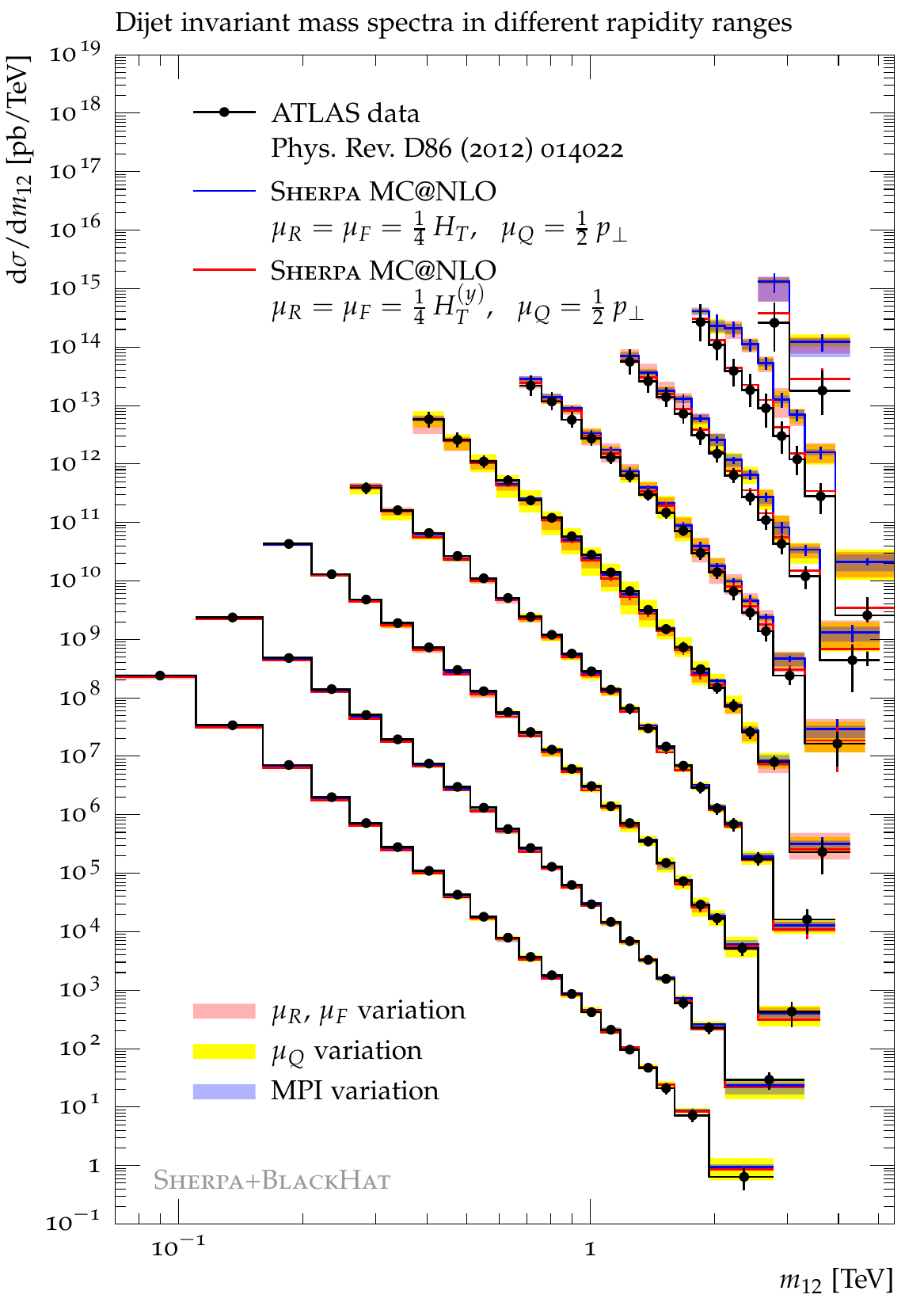}
  \end{minipage}
  \hfill
  \begin{minipage}{0.47\textwidth}
    \lineskip-1.85pt
    \includegraphics[width=\textwidth]{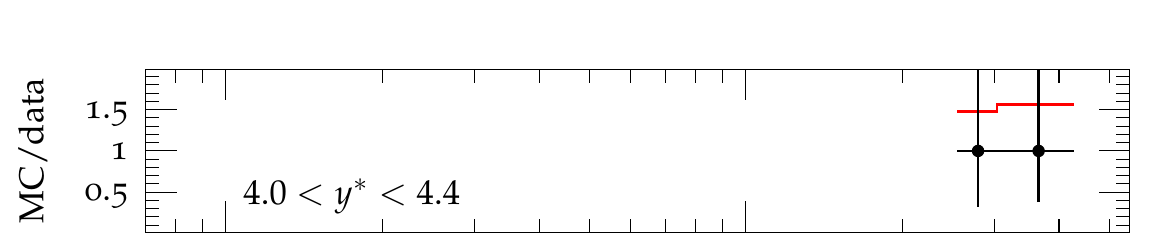}\\
    \includegraphics[width=\textwidth]{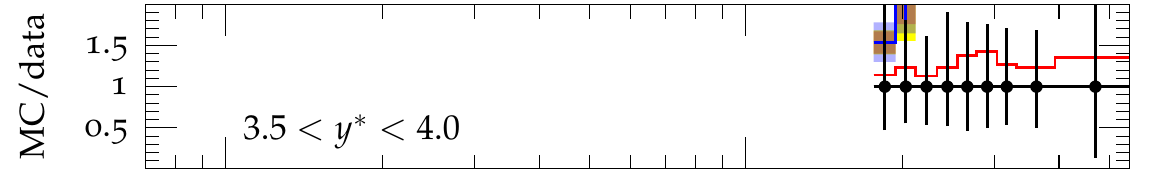}\\
    \includegraphics[width=\textwidth]{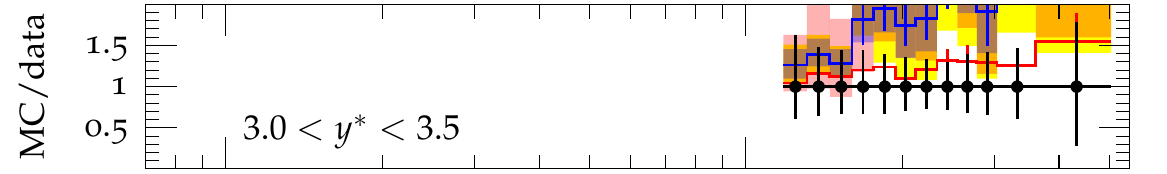}\\
    \includegraphics[width=\textwidth]{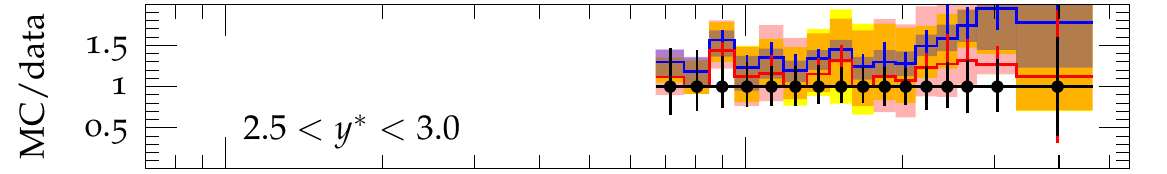}\\
    \includegraphics[width=\textwidth]{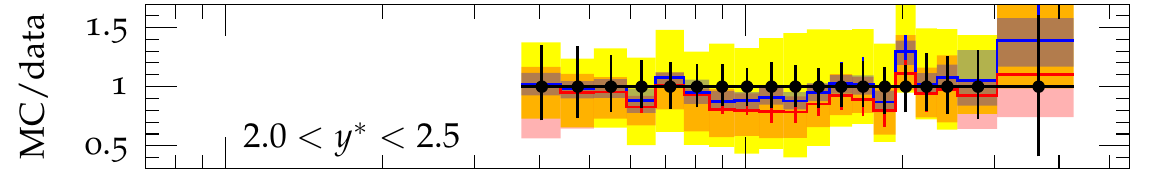}\\
    \includegraphics[width=\textwidth]{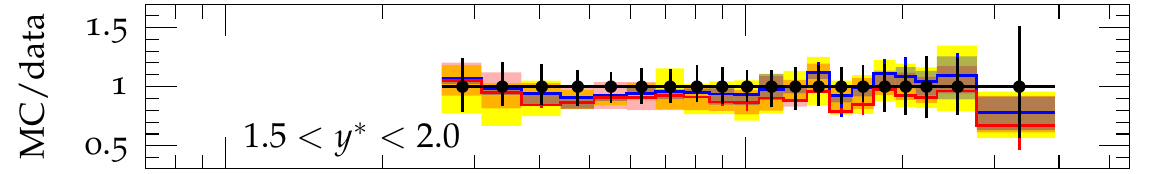}\\
    \includegraphics[width=\textwidth]{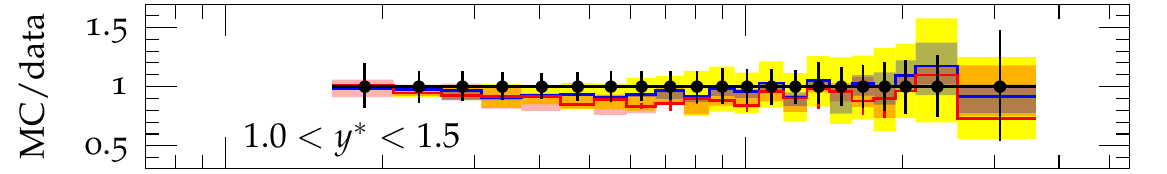}\\
    \includegraphics[width=\textwidth]{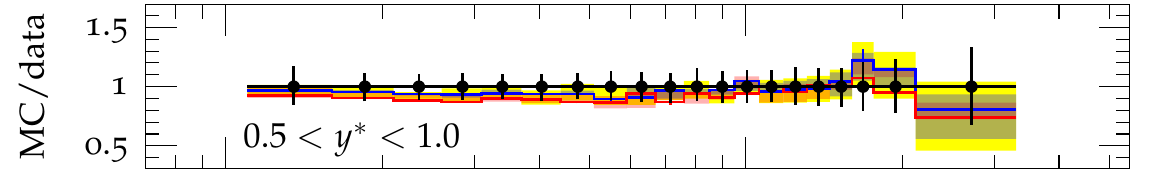}\\
    \includegraphics[width=\textwidth]{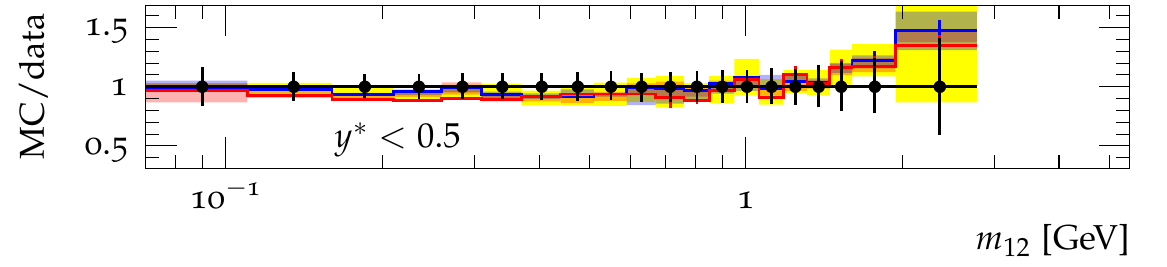}
  \end{minipage}
  \caption{
           Dijet invariant masses in different rapidity ranges compared to 
           \ATLAS data \cite{Aad:2011fc}. The rapidity different difference 
           ranges on the left-hand side plot are ordered as indicated in the 
           right-hand side ratio plots, being multiplied, from top to bottom 
           by factors of $10^{12}$, $3\!\cdot\!10^{10}$, $10^{9}$, 
           $3\!\cdot\!10^{8}$, $10^{6}$, $3\!\cdot\!10^{4}$,  $10^{3}$, 
           $30$, and $1$, for readabilities sake. 
           \label{Fig:ATLAS_dijet_masses}
          }
\end{figure}

\subsection*{Dijet invariant masses in different rapidity ranges}
Another doubly differential observable studied in \cite{Aad:2011fc} is the 
dijet invariant mass. Events with at least two anti-$k_\perp$ jets with 
$R=0.4$ and $p_\perp^\text{lead}>30$ GeV and  $p_\perp^\text{sublead}>20$ 
GeV within $|y|<4.4$ are considered. The dijet invariant mass is defined
$m_{12}=\sqrt{(p^\text{lead}+p^\text{sublead})^2}$ and binned into various 
mutually exclusive ranges of 
$y^*=\tfrac{1}{2}\,|y_\text{lead}-y_\text{sublead}|$, the rapidity separation 
of the two jets. Fig.\ \ref{Fig:ATLAS_dijet_masses} displays MC results 
compared to \ATLAS data. For small rapidity separations the agreement with 
data is good. At large $y^*$ the renormalisation and factorisation scale is 
too low, resulting in too large MC predictions. This situation is improved 
by choosing the scale \EqRef{eq:hty} instead. The renormalisation and resummation 
scale uncertainties are small at small $y^*$ and increase towards larger $y^*$. 
They also remain approximately constant over the whole considered $m_{12}$ 
range. Resummation scale uncertainties, on the other hand, are larger 
and display a definite $m_{12}$ dependence. This can be seen as an indication
that high-energy resummation, which goes beyond the collinear limit used in the 
parton shower, becomes important~\cite{Andersen:2009nu,Andersen:2009he,Andersen:2011hs}.
Non-perturbative uncertainties 
are only non-negligible at low invariant masses or large rapidity 
separations, when at least one jet is likely to be close to either 
of the two beams.

As discussed in the previous paragraph, taking $\tfrac{1}{4}\,H_T^{(y)}$ as 
the central scale improves the description of this observable but 
leads to a worse description of the other observables investigated in this 
paper. It is therefore not adopted as the central scale.

\begin{figure}[t!]
  \begin{minipage}{0.47\textwidth}
    \includegraphics[width=\textwidth]{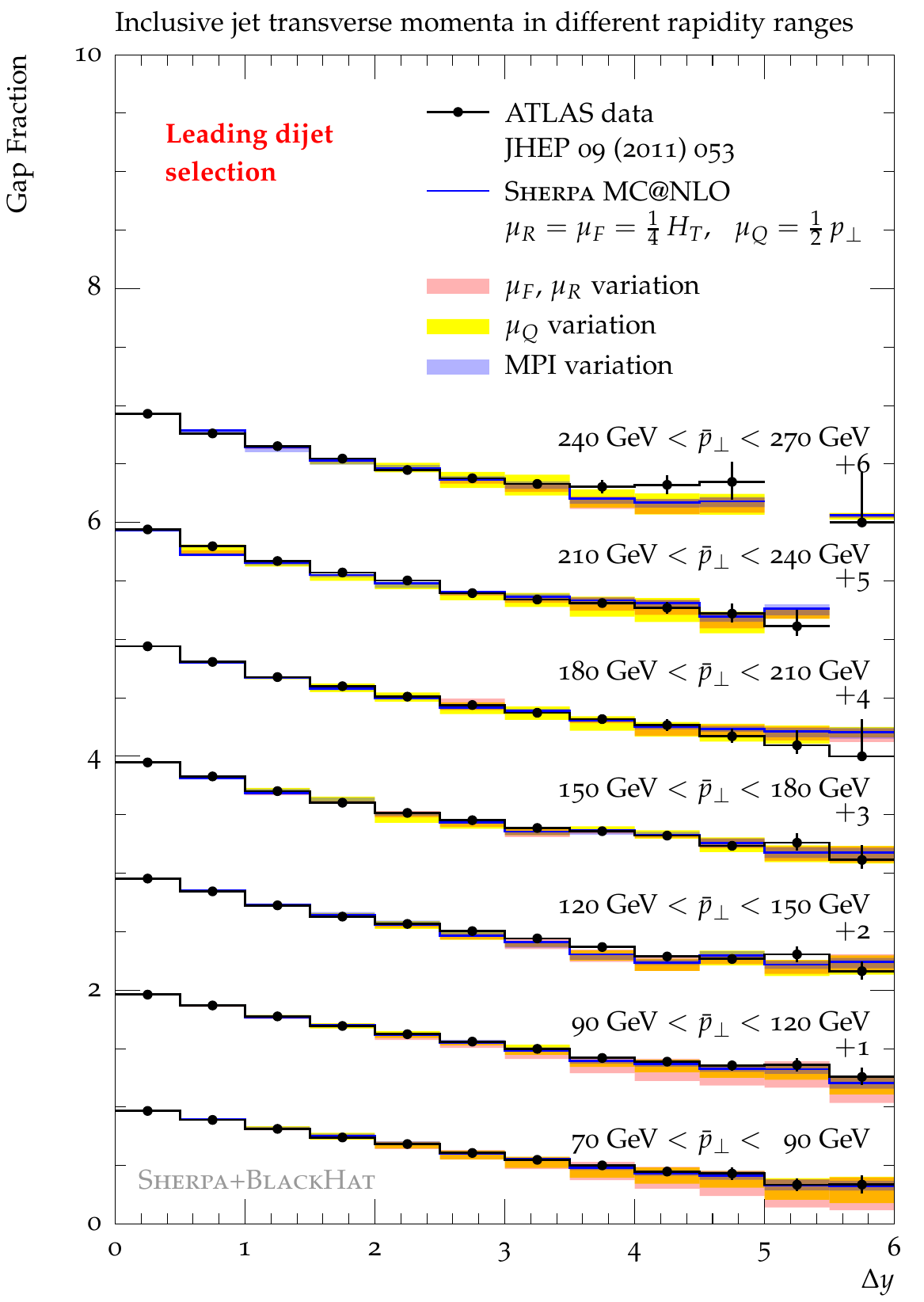}
  \end{minipage}
  \hfill
  \begin{minipage}{0.47\textwidth}
    \lineskip-1.85pt
    \includegraphics[width=\textwidth]{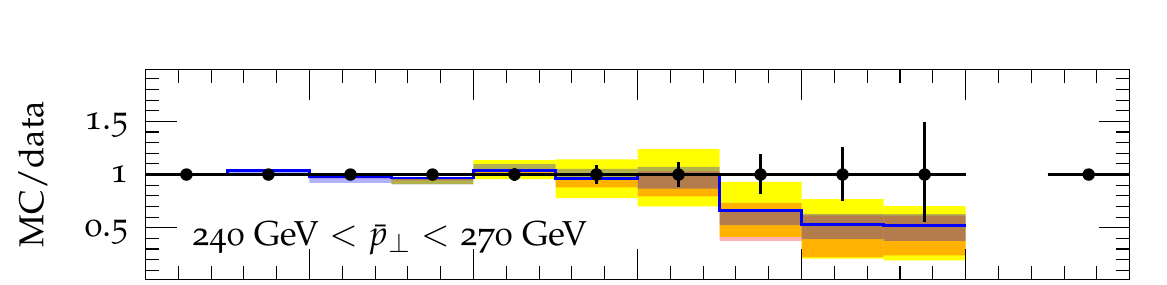}\\
    \includegraphics[width=\textwidth]{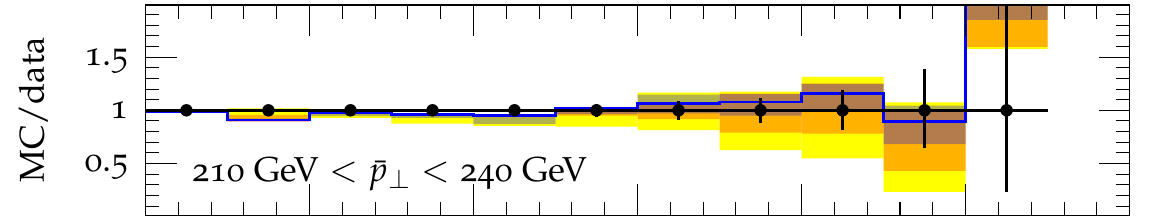}\\
    \includegraphics[width=\textwidth]{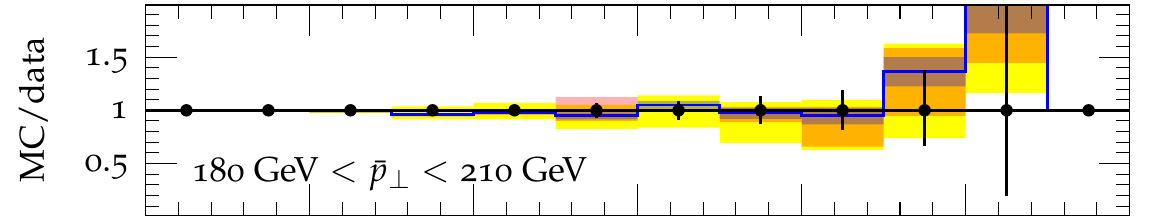}\\
    \includegraphics[width=\textwidth]{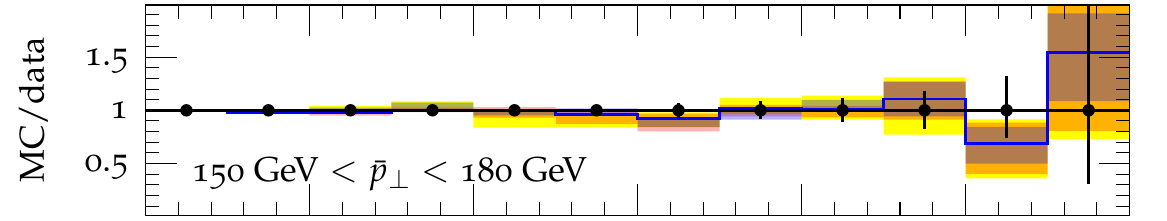}\\
    \includegraphics[width=\textwidth]{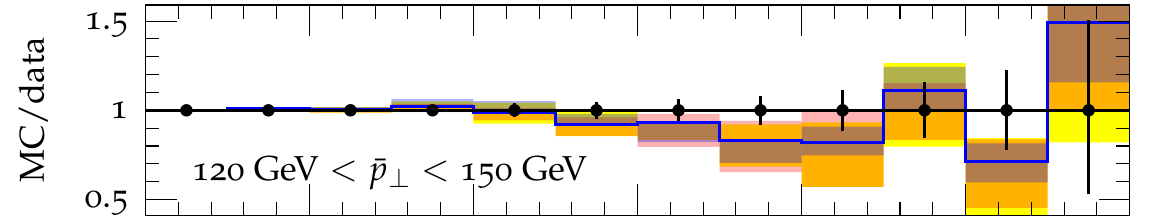}\\
    \includegraphics[width=\textwidth]{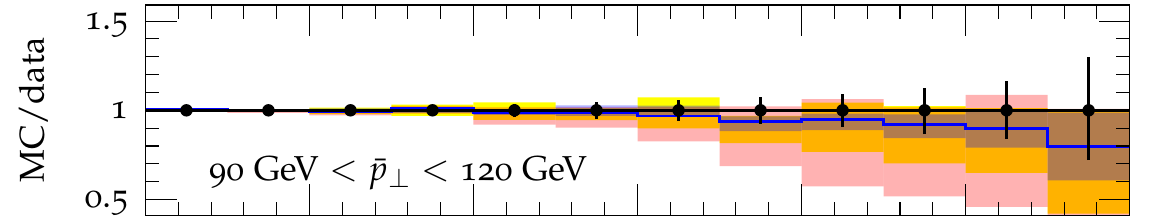}\\
    \includegraphics[width=\textwidth]{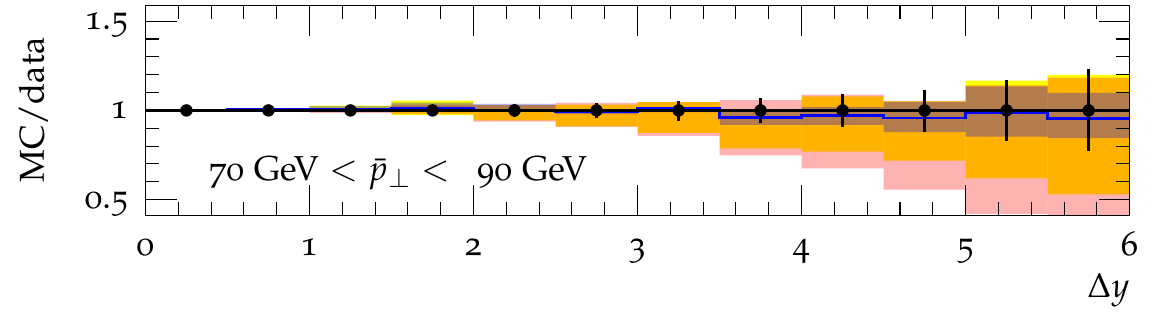}
  \end{minipage}
  \vspace*{2mm}\\
  \begin{minipage}{0.47\textwidth}
    \includegraphics[width=\textwidth]{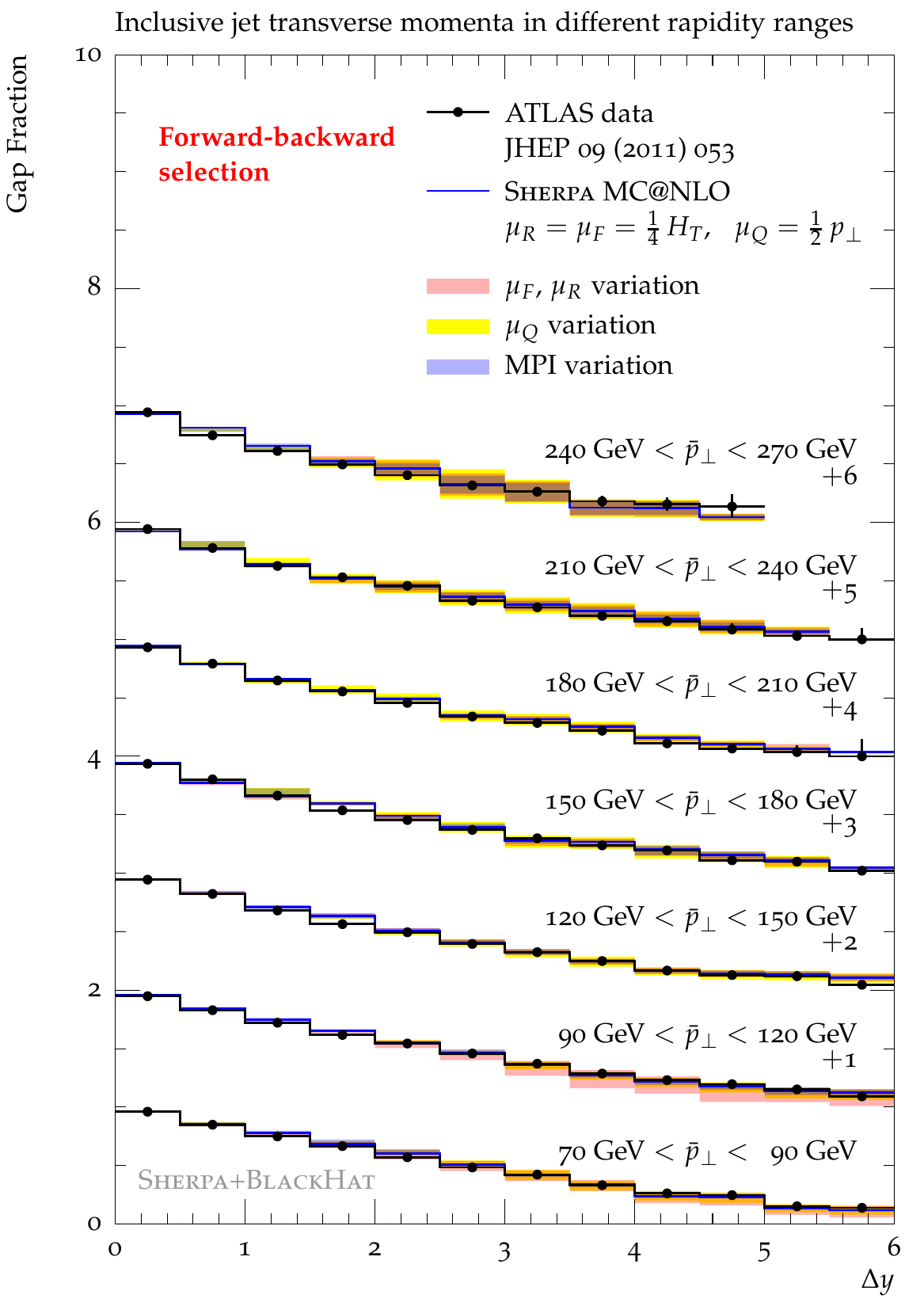}
  \end{minipage}
  \hfill
  \begin{minipage}{0.47\textwidth}
    \lineskip-1.85pt
    \includegraphics[width=\textwidth]{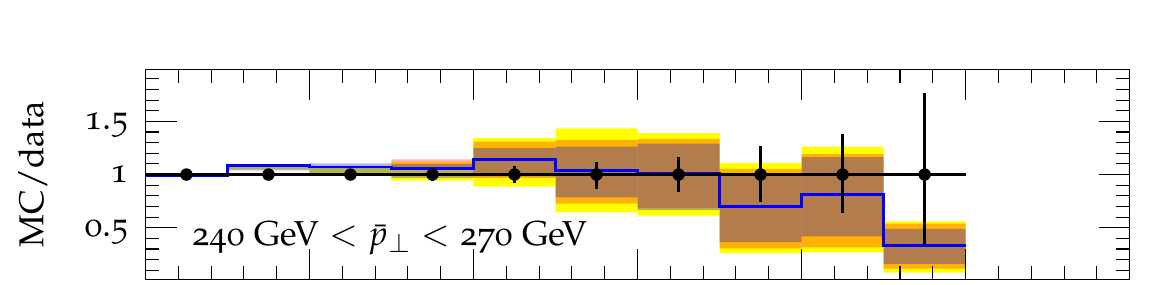}\\
    \includegraphics[width=\textwidth]{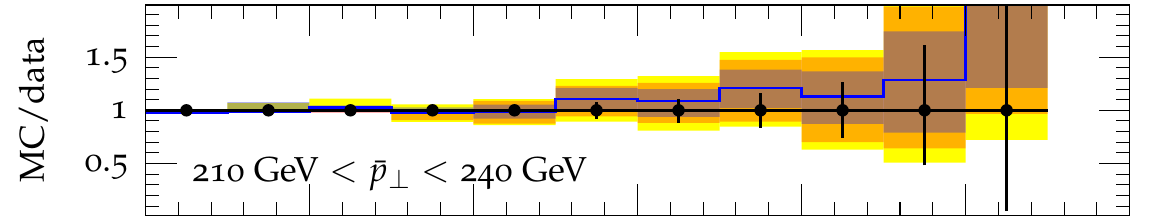}\\
    \includegraphics[width=\textwidth]{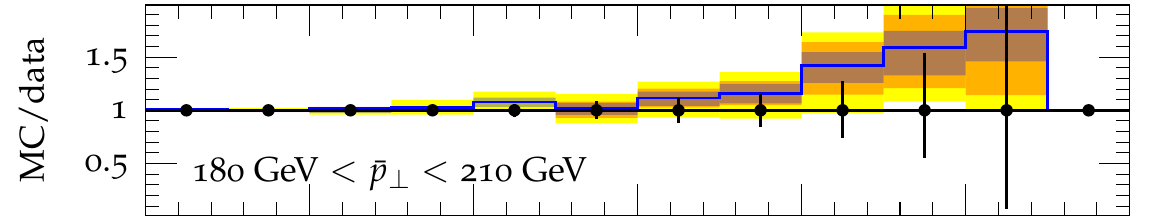}\\
    \includegraphics[width=\textwidth]{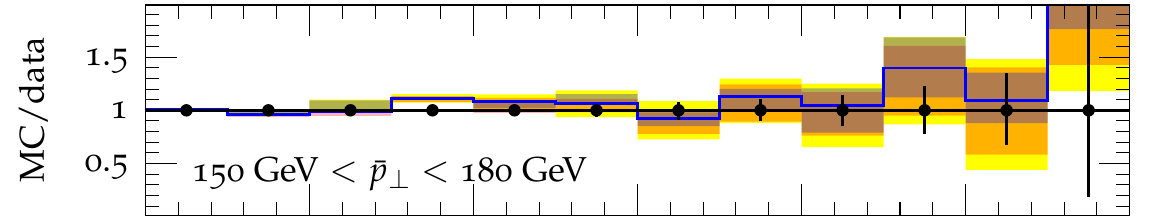}\\
    \includegraphics[width=\textwidth]{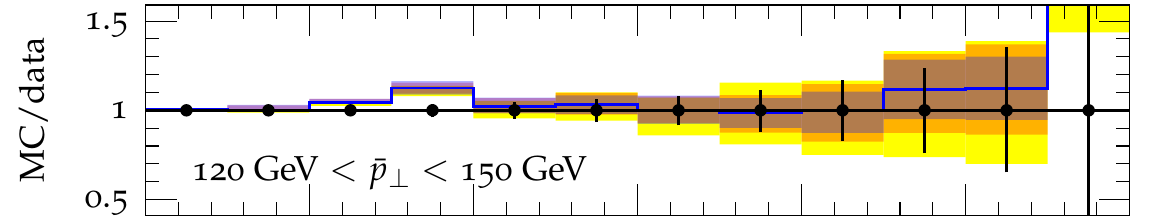}\\
    \includegraphics[width=\textwidth]{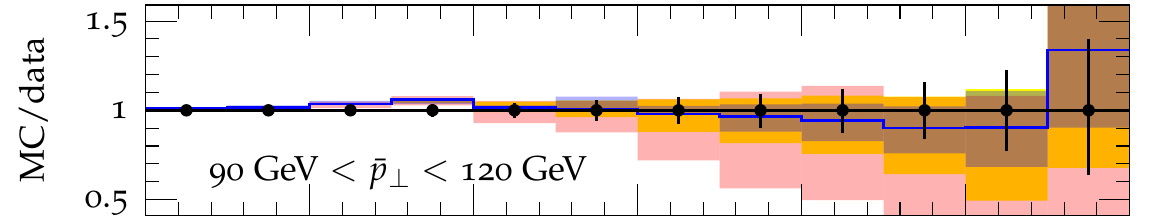}\\
    \includegraphics[width=\textwidth]{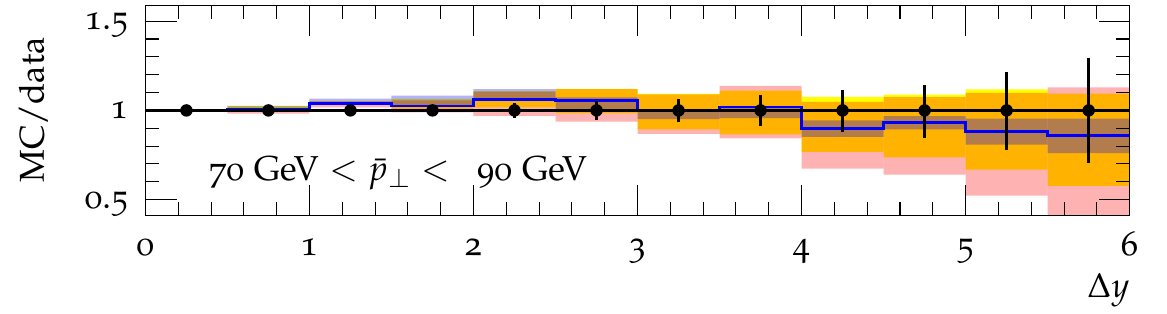}
  \end{minipage}
  \caption{
           Gap fraction in dependence of mean transverse momentum and 
           rapidity separation of dijet pair for both selections 
           compared to \ATLAS data \cite{Aad:2011jz}.
           \label{Fig:ATLAS_Gapfraction_dy_AB}
          }
\end{figure}

\begin{figure}[t!]
  \begin{minipage}{0.47\textwidth}
    \includegraphics[width=\textwidth]{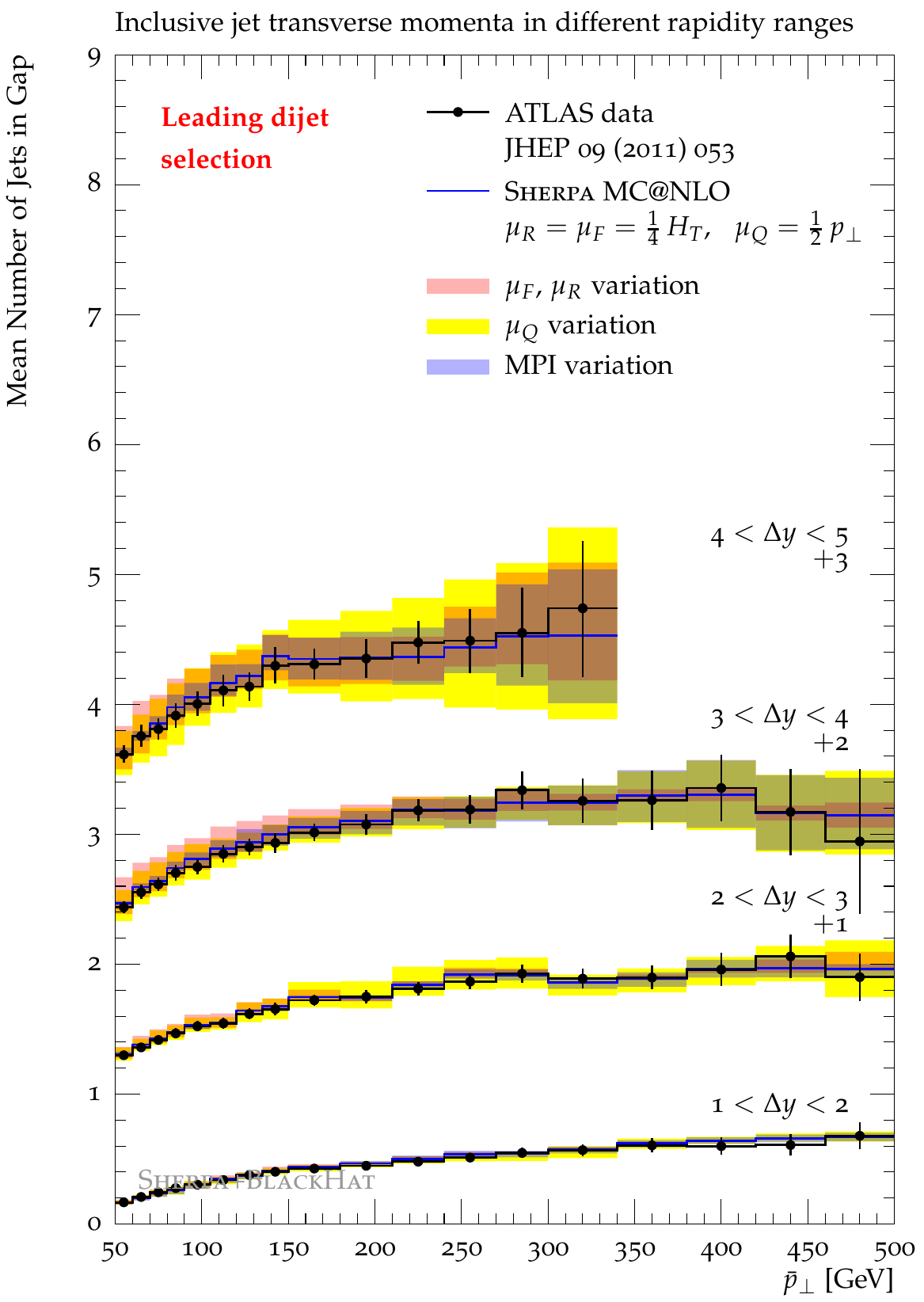}
  \end{minipage}
  \hfill
  \begin{minipage}{0.47\textwidth}
    \lineskip-1.85pt
    \includegraphics[width=\textwidth]{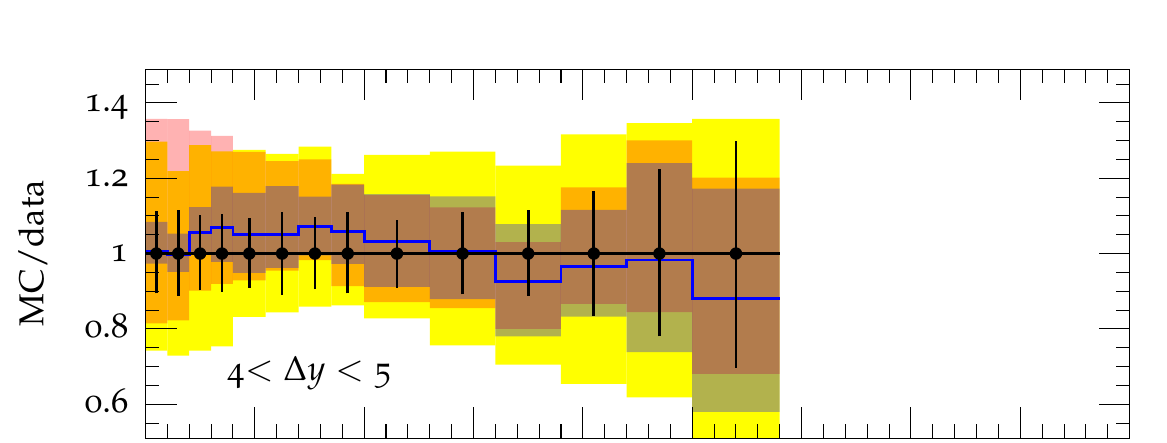}\\
    \includegraphics[width=\textwidth]{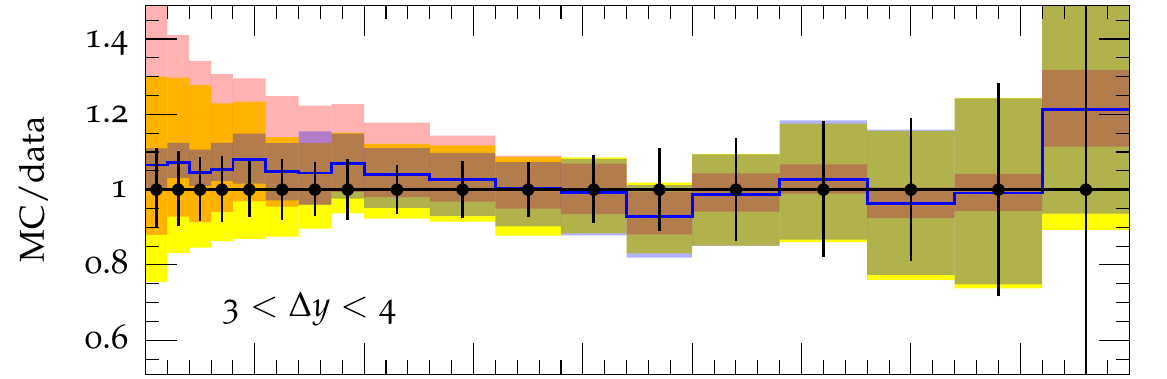}\\
    \includegraphics[width=\textwidth]{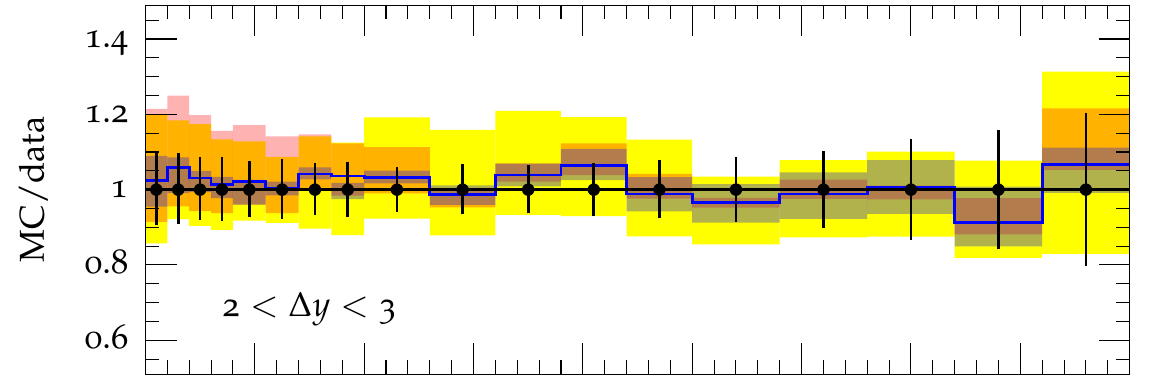}\\
    \includegraphics[width=\textwidth]{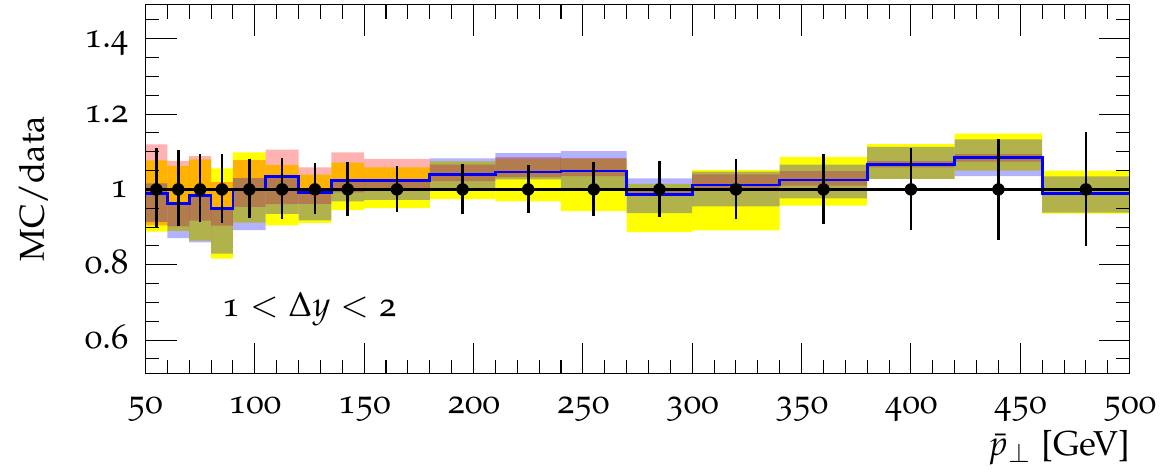}
  \end{minipage}
  \vspace*{2mm}\\
  \begin{minipage}{0.47\textwidth}
    \includegraphics[width=\textwidth]{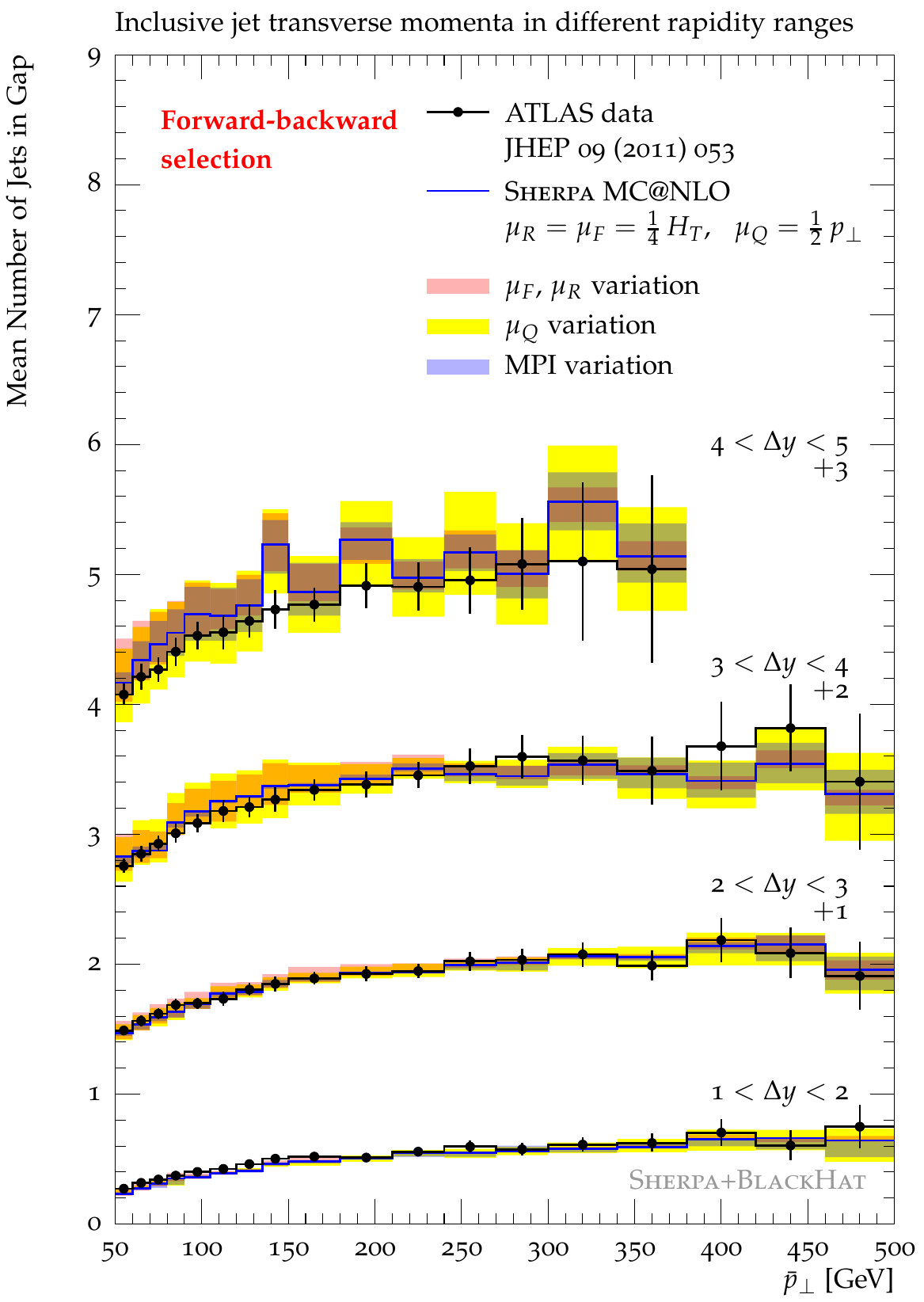}
  \end{minipage}
  \hfill
  \begin{minipage}{0.47\textwidth}
    \lineskip-1.85pt
    \includegraphics[width=\textwidth]{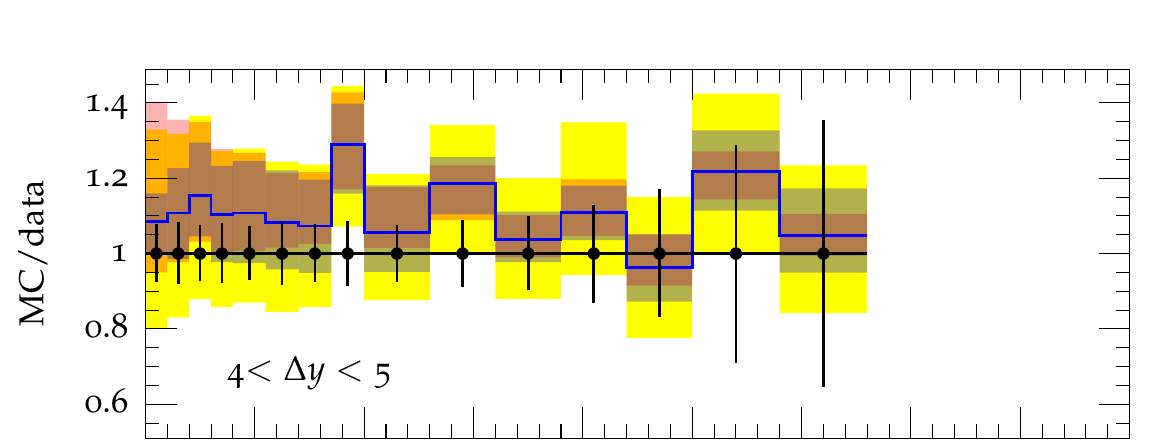}\\
    \includegraphics[width=\textwidth]{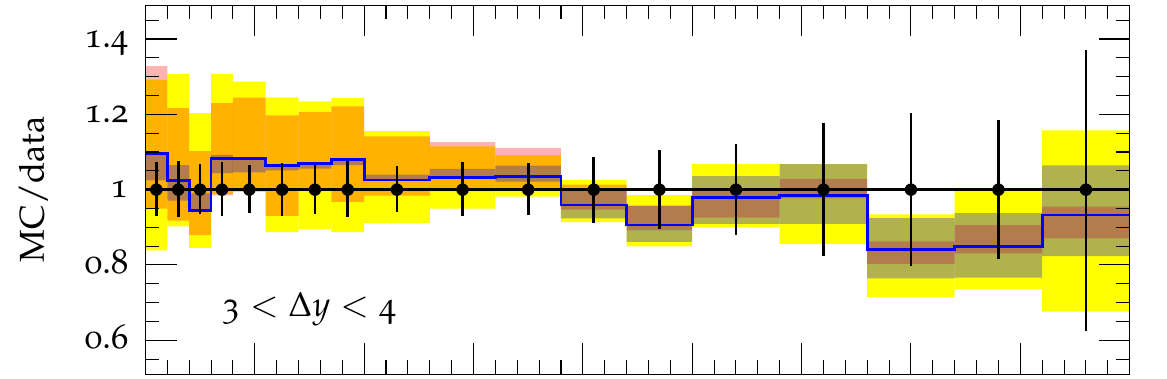}\\
    \includegraphics[width=\textwidth]{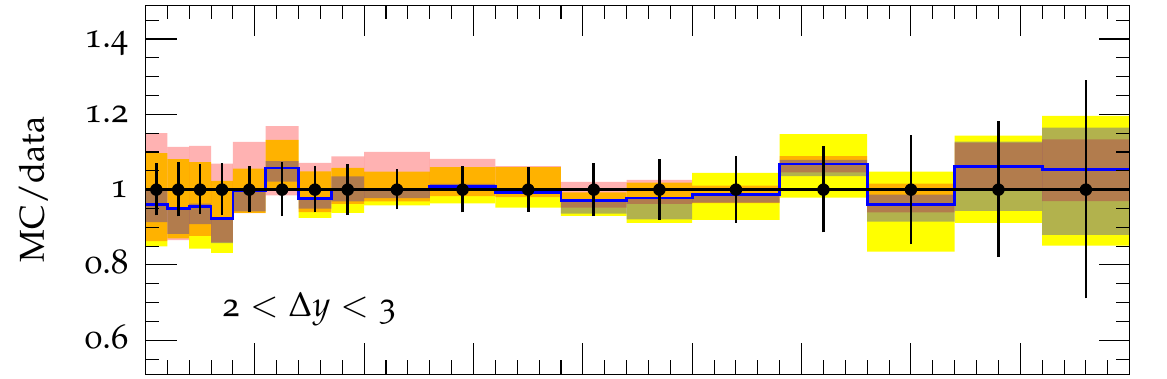}\\
    \includegraphics[width=\textwidth]{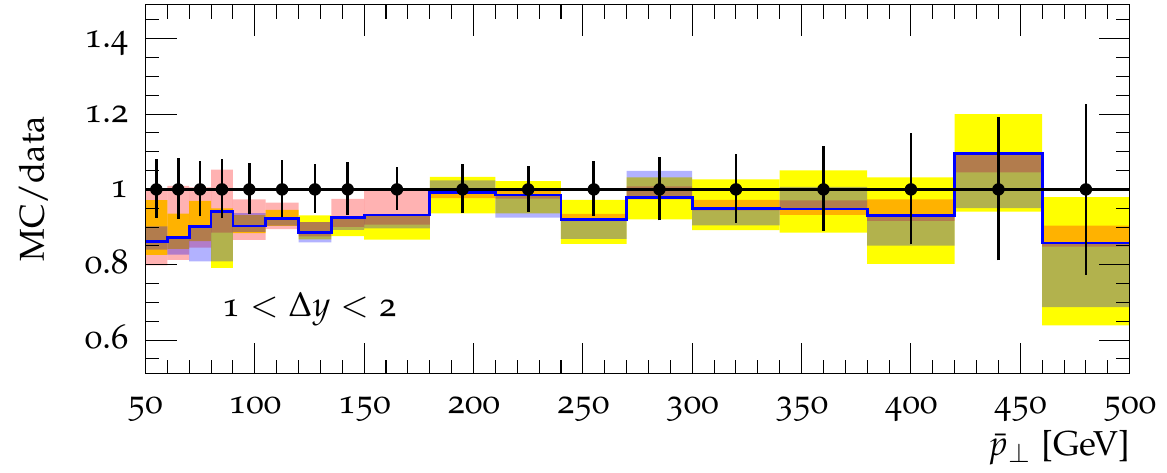}
  \end{minipage}
  \caption{
           Mean jet number in gap  in dependence of mean transverse 
           momentum and rapidity separation of dijet pair for both 
           selections compared to \ATLAS data \cite{Aad:2011jz}.
           \label{Fig:ATLAS_NJet_pT_AB}
          }
\end{figure}

\subsection*{Gap fractions}
A different way to probe the radiation pattern was explored by the 
\ATLAS collaboration in \cite{Aad:2011jz}. Therein, events were selected 
containing at least two jets, defined using the anti-$k_\perp$ algorithm 
with $R=0.6$, each with $p_\perp>20$ GeV within $y<4.4$. Within these 
events a dijet system is then identified using either the two largest 
transverse momentum jets (leading jet selection) or the widest separated 
jets (forward backward selection). For both definitions an average 
transverse momentum $\bar{p}_\perp=\tfrac{1}{2}(p_\perp^\text{jet1}+p_\perp^\text{jet2})$ 
of at least $50$ GeV is required. To characterise the subsequent radiation 
pattern two variables are used. The gap fraction, i.e.\ the fraction of 
events that do not exhibit any further radiation above some $Q_0$ within 
the $\Delta y$ rapidity range spanned by the dijet system, and the mean
number of jets with $p_\perp>Q_0$ in the same $\Delta y$ region. Fig.\ 
\ref{Fig:ATLAS_Gapfraction_dy_AB} 
displays a comparison of MC results with data for the gap fraction ($Q_0=20$ GeV) in dependence 
on the rapidity separation of the dijet system and its $\bar{p}_\perp$ for 
both selections. The agreement is good throughout the probed region. In 
both selections the renormalisation and factorisation scale uncertainty 
is the leading uncertainty at low $\bar{p}_\perp$ while the resummation 
uncertainty dominates at large $\bar{p}_\perp$. Non-perturbative 
uncertainties are generally larger than for most other observables 
considered in this paper, increasing with $\Delta y$ and $\bar{p}_\perp$. 
In case of the forward backward selection they are of the same magnitude 
as both perturbative uncertainties in the large $\Delta y$ and 
$\bar{p}_\perp$ region.

Similarly, Fig.~\ref{Fig:ATLAS_NJet_pT_AB} displays the average number of 
jets ($Q_0=20$ GeV) in dependence on the rapidity separation of the dijet 
system and its $\bar{p}_\perp$ for both selections. Again, good agreement 
is found. All uncertainties are small and of comparable size for small 
$\Delta y$ throughout the $\bar{p}_\perp$ range, and steadily increasing 
for larger $\Delta y$. While a resummation scale variation produces 
largely $\bar{p}_\perp$ independent uncertainties the renormalisation and 
factorisation scale uncertainties are larger for small average transverse 
momenta than for large ones. The non-perturbative uncertainties show the 
opposite behaviour.

\begin{figure}[t!]
  \begin{minipage}{0.47\textwidth}
    \includegraphics[width=\textwidth]{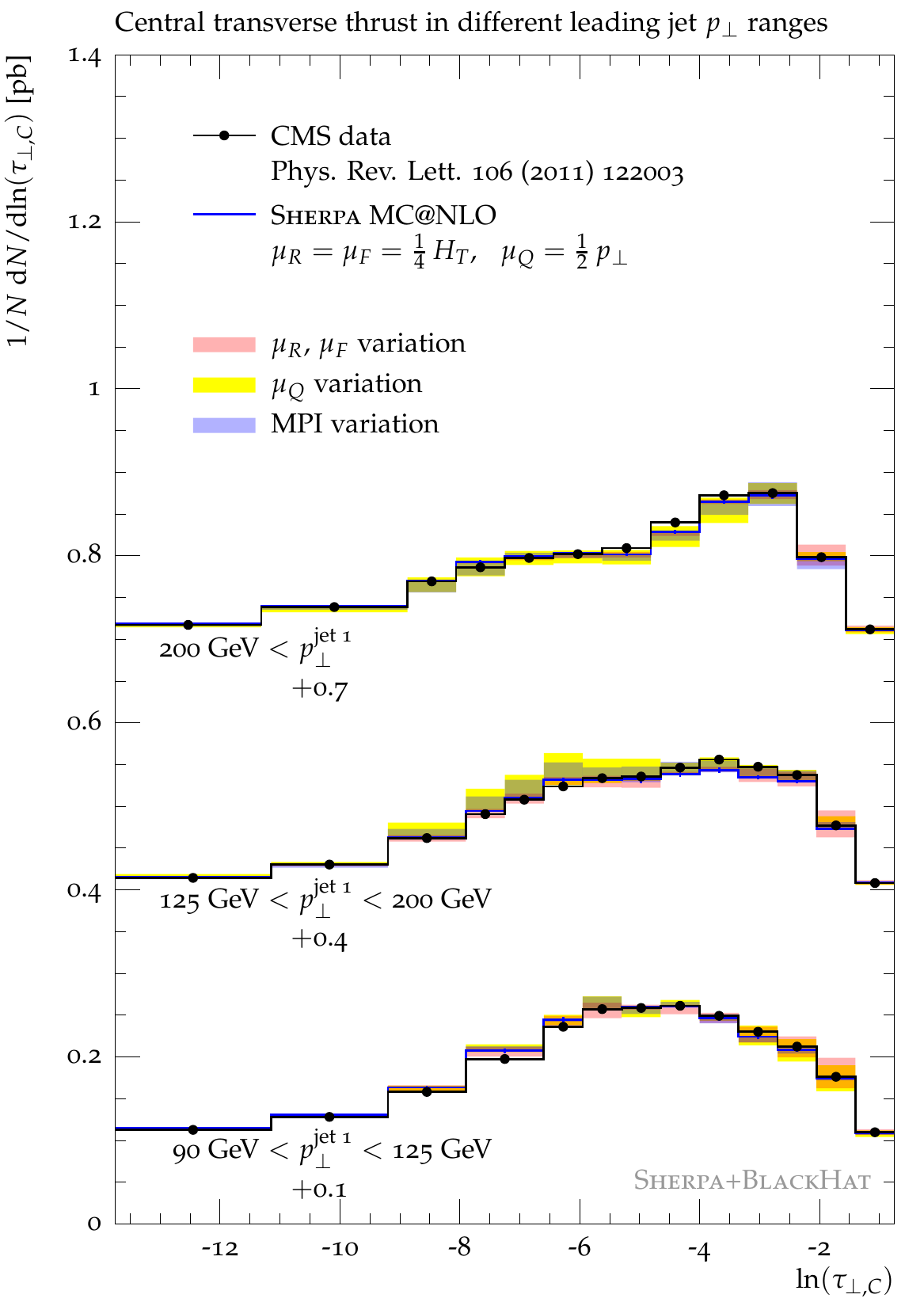}
  \end{minipage}
  \hfill
  \begin{minipage}{0.47\textwidth}
    \lineskip-1.85pt
    \includegraphics[width=\textwidth]{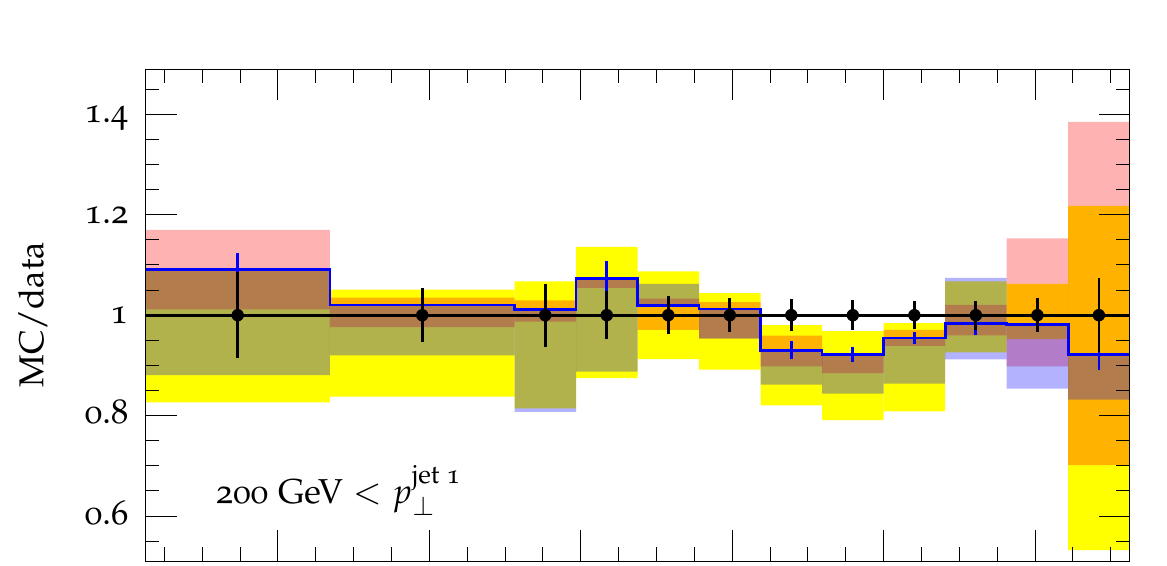}\\
    \includegraphics[width=\textwidth]{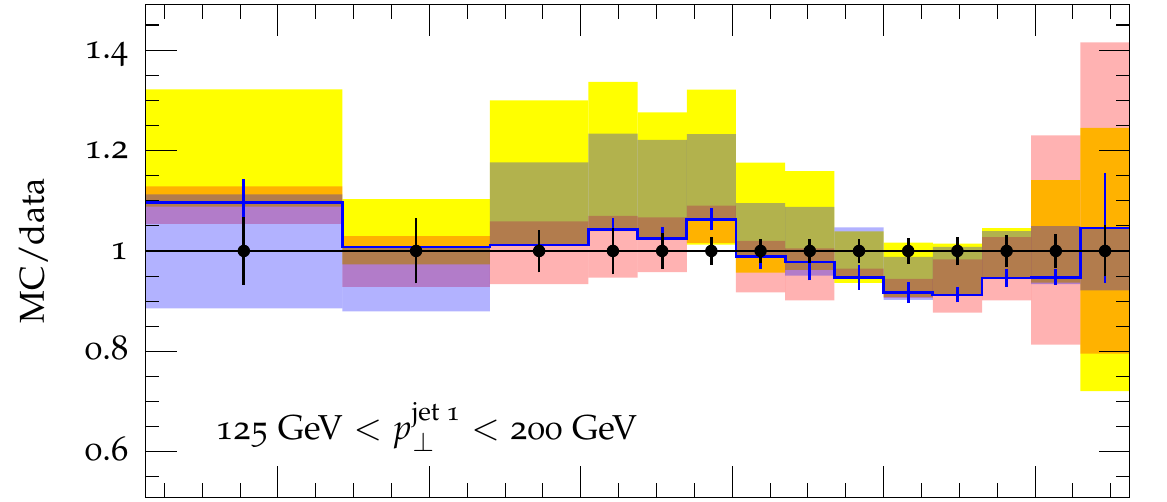}\\
    \includegraphics[width=\textwidth]{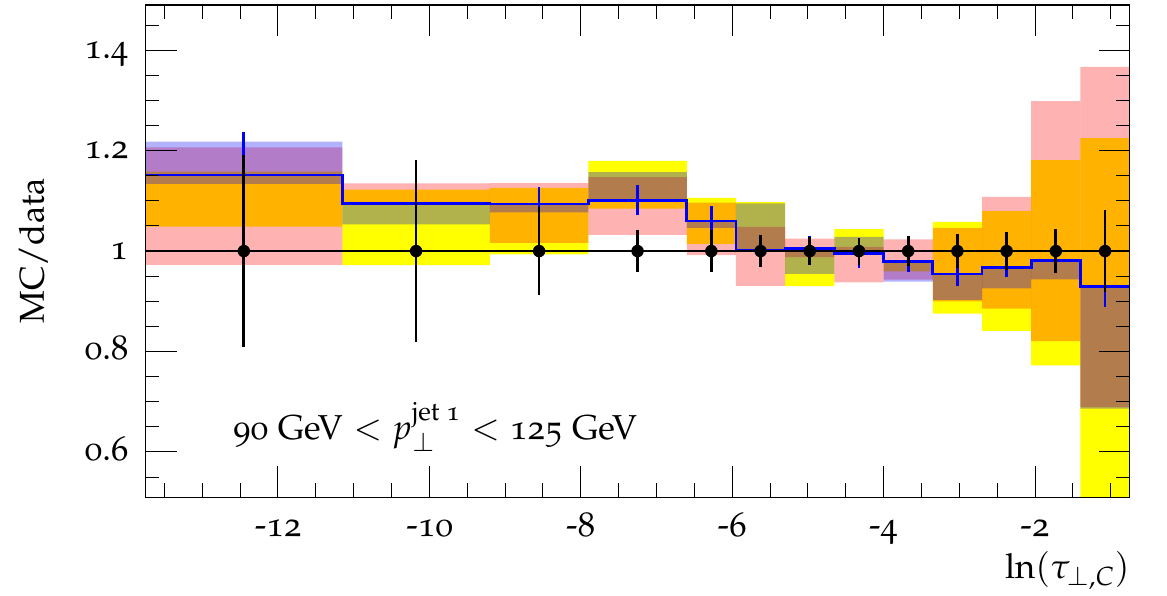}
  \end{minipage}
  \vspace*{2mm}\\\noindent
  \begin{minipage}{0.47\textwidth}
    \includegraphics[width=\textwidth]{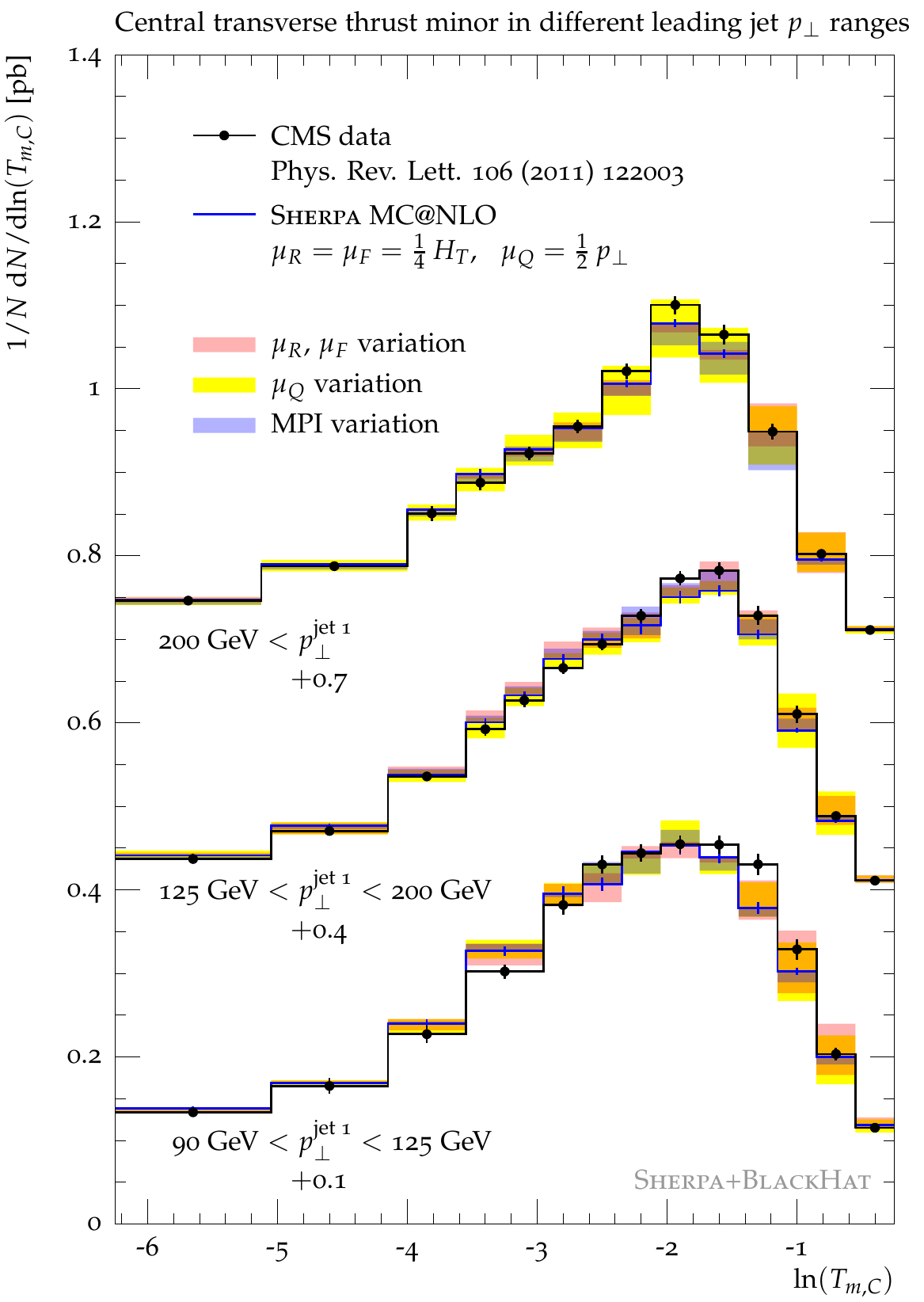}
  \end{minipage}
  \hfill
  \begin{minipage}{0.47\textwidth}
    \lineskip-1.85pt
    \includegraphics[width=\textwidth]{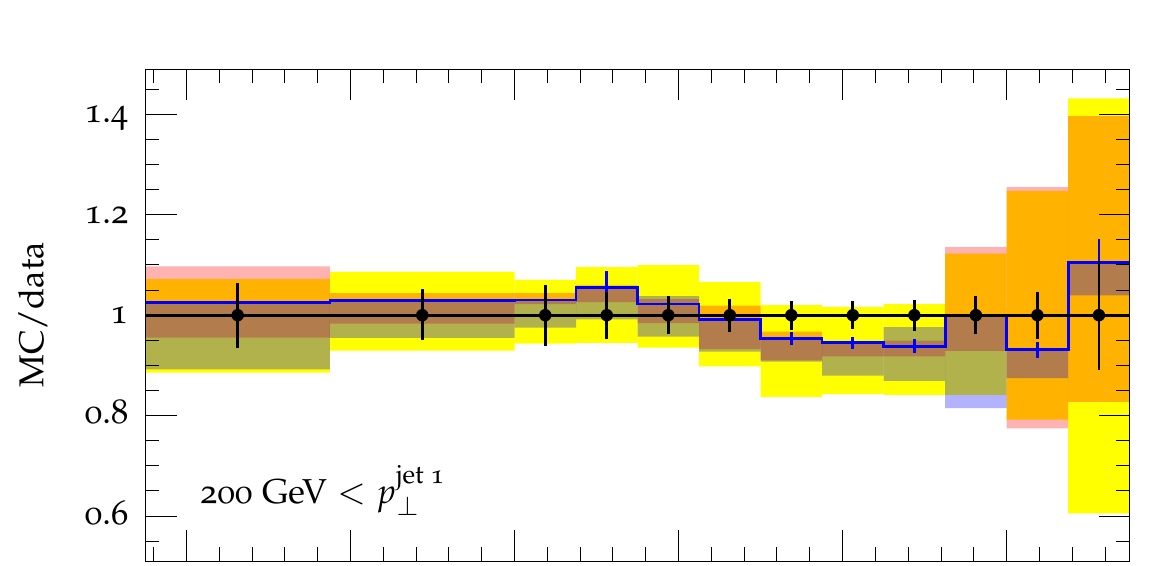}\\
    \includegraphics[width=\textwidth]{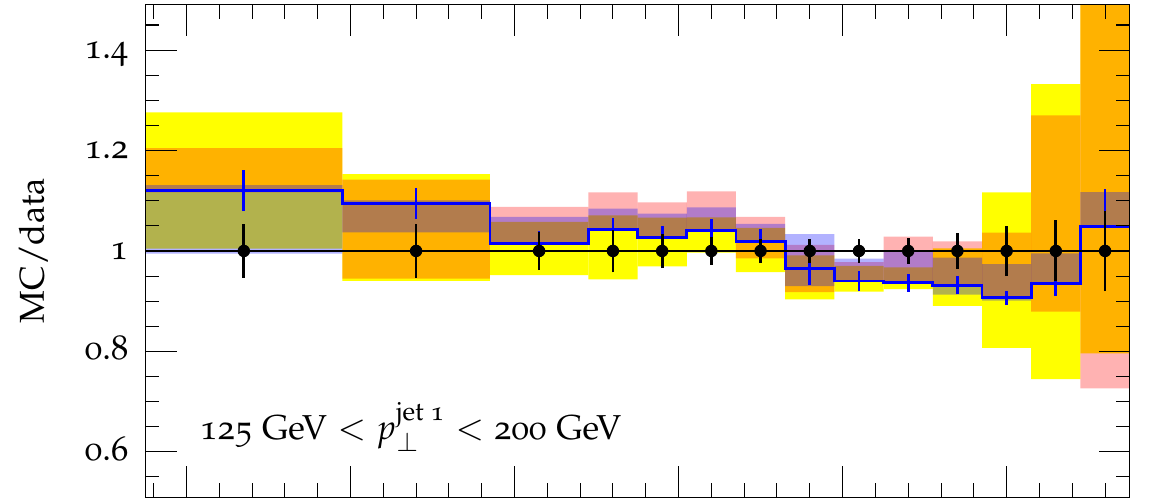}\\
    \includegraphics[width=\textwidth]{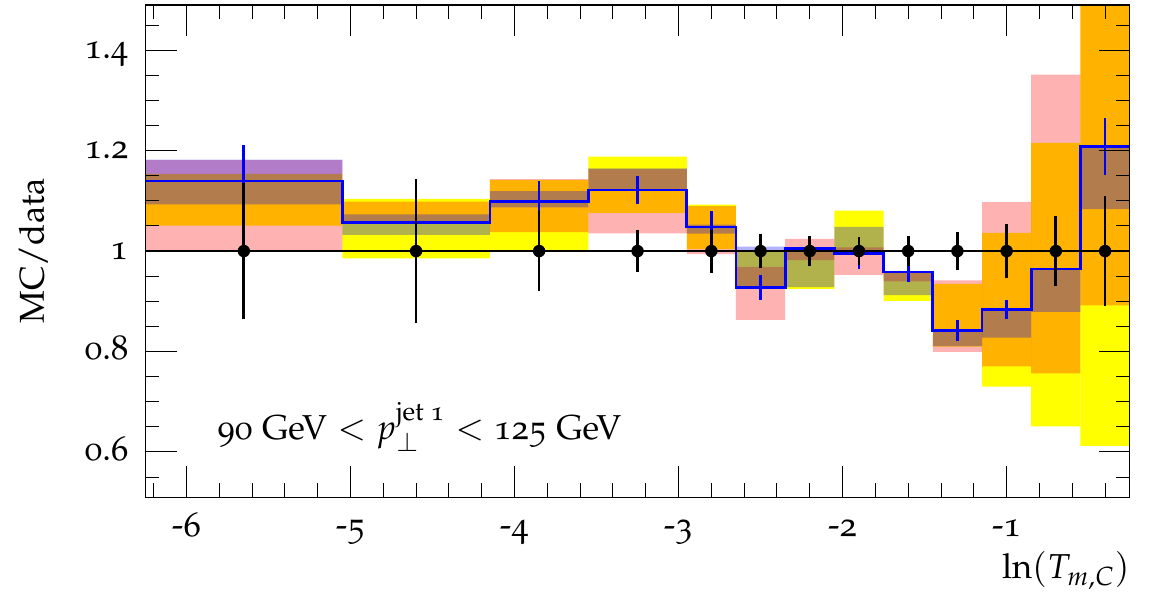}
  \end{minipage}
  \caption{
           Central transverse thrust and central transverse thrust minor 
           compared to \CMS data \cite{Khachatryan:2011dx}.
           \label{Fig:CMS_thrust}
          }
\end{figure}

\subsection*{Event shapes}
Traditional observables not being described without resummation 
are event shapes. The \CMS collaboration measured 
the central transverse thrust and the central transverse thrust minor in 
multijet production in \cite{Khachatryan:2011dx}. The sample is defined 
by requiring at least two jets, defined using the anti-$k_\perp$ algorithm 
with $R=0.5$, $p_\perp>30$~GeV and $|\eta|<1.3$. The selected events are then 
categorised into three mutually exclusive regions according to the 
leading jet transverse momentum. The observables are defined as 
follows
\begin{equation}
  \tau_{\perp,C} = 1-\max\limits_{\hat n_T}
                   \frac{\sum_{i\in\text{jets}}|\vec{p}_{\perp,i}\cdot\hat n_T|}
                        {\sum_{i\in\text{jets}}|\vec{p}_{\perp,i}|}
  \qquad\qquad\text{and}\qquad\qquad
  T_{m,C} = \frac{\sum_{i\in\text{jets}}|\vec{p}_{\perp,i}\times\hat n_{T,C}|}
                 {\sum_{i\in\text{jets}}|\vec{p}_{\perp,i}|}\;.\nnb
\end{equation}
Therein, the vector $n_{T,C}$ is defined as the vector minimising 
$\tau_{\perp,C}$. Only jet momenta are taken into account. The results of 
the presented calculation are compared to the experimental data in 
Fig.~\ref{Fig:CMS_thrust}. Good agreement between MC predictions and data is found. The dependence of 
both observables on the renormalisation and factorisation scale, despite
being calculated at most at leading order accuracy, largely cancels due to 
their normalisation. However, they 
do show a large dependence on the resummation scale, as expected. This is 
largest when $\tau_{\perp,C}$ ($T_{m,C}$) is close to zero (one). The 
non-perturbative uncertainties have the opposite behaviour.

\begin{figure}[t]
  \centering
  \includegraphics[width=0.47\textwidth]{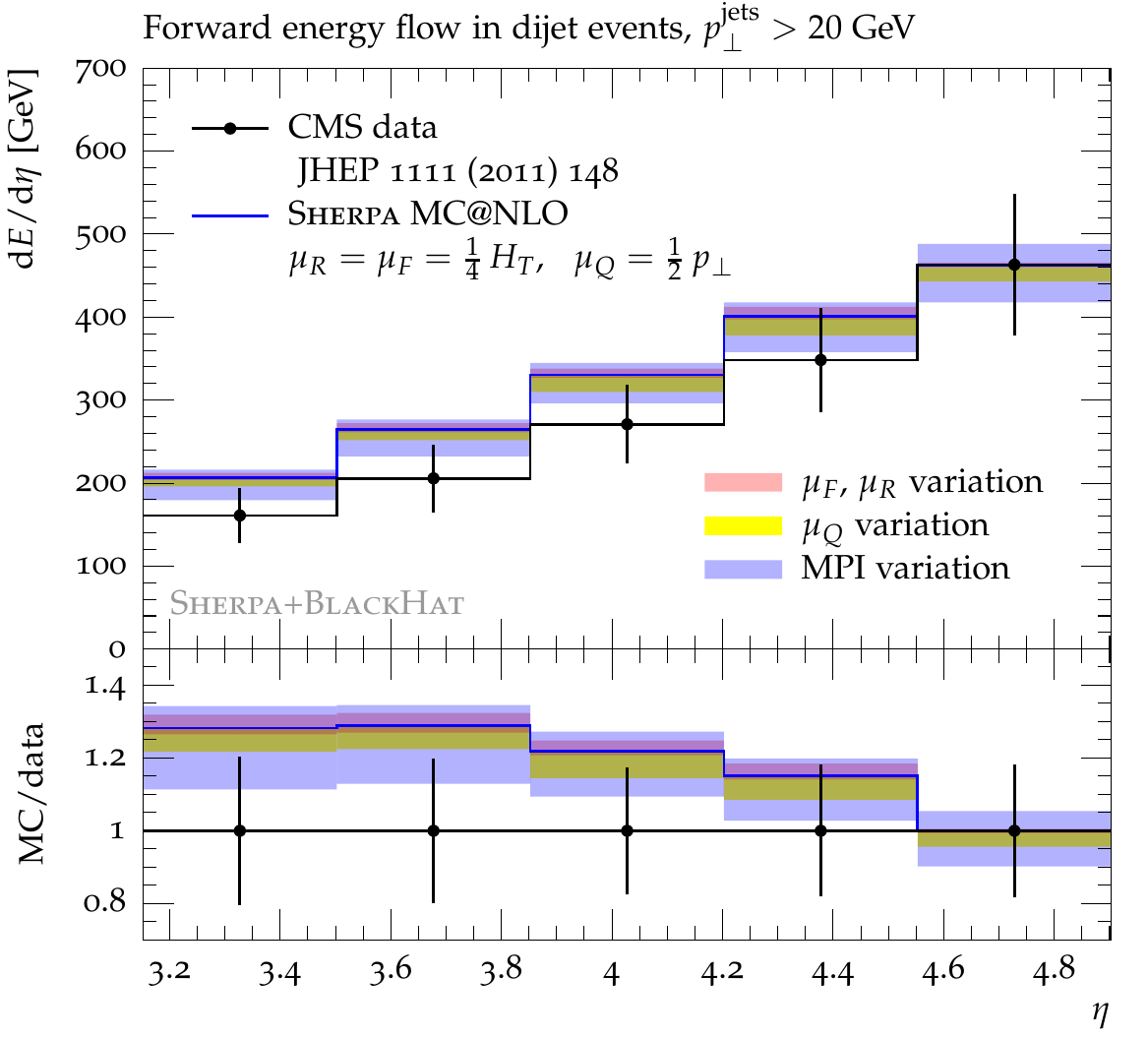}\hfill
  \caption{
           Forward energy flow compared to \CMS data \cite{Chatrchyan:2011wm}.
           \label{Fig:CMS_forward_enery_flow}
          }
\end{figure}

\subsection*{Forward energy flow}
The last observable to be studied is the forward energy flow as measured by 
the \CMS collaboration \cite{Chatrchyan:2011wm}. In this analysis, events 
with at least two jets, defined by the anti-$k_\perp$ algorithm with $R=0.5$ 
and $p_\perp>20$ GeV are required. The two leading jets are further required 
to lie within $|\eta|<2.5$ and satisfy $|\Delta\phi-\pi|<1$, i.e. to produce a 
nearly back-to-back topology. Within this event sample the energy 
flow, defined as average energy $E$ per event per pseudo-rapidity interval 
$\done\eta$, is calculated and compared to the measured data. Despite 
a small difference in shape good agreement is found. 
Again, due to the normalisation of the observable, renormalisation and 
factorisation scale uncertainties as well as resummation scale uncertainties 
are small. A comparably large uncertainty stems from the non-perturbative 
modelling uncertainties, ranging up to $\sim$10\%. This is not unexpected 
since in this very forward region, close to the beams, non-factorizable
components of the inclusive cross section play a large role.

\section{Conclusions}
\label{Sec:Conclusions}

We have presented a detailed analysis of uncertainties associated 
with the simulation of inclusive jet and dijet production using methods
for matching next-to-leading order QCD calculations and parton showers.
We have analysed factorisation and renormalisation scale dependence
as well as variations originating from the choice of resummation scale.
We have compared to uncertainties originating from the freedom in choosing 
parameters in the Monte-Carlo simulation of multiple parton scattering.

These three types of uncertainties represent different degrees of freedom
in the Monte-Carlo simulation: While the renormalisation and factorisation
scale dependence probe the impact of higher-order QCD corrections to the 
hard process, the resummation scale dependence quantifies, to some extent, 
uncertainties related to parton evolution. Variations of the MPI tune
are used to estimate uncertainties related to non-perturbative dynamics.

The results of our Monte-Carlo simulation have been compared to a variety 
of data taken by the \ATLAS and \CMS experiments at the CERN \LHC. 
Good agreement is found for almost all observables. Exceptions are 
the inclusive jet transverse momenta at large jet rapidity and the dijet 
invariant masses at large average rapidity. Discrepancies are attributed 
to the choice of renormalisation and factorisation scale, 
which is given by one quarter of the visible transverse energy. Despite 
taking into account real-emission dynamics, this scale does not give 
a realistic measure of the hardness of events at large individual 
jet rapidities. A modified scale, with jet transverse momenta weighted by
their rapidity w.r.t.\ the centre of the partonic system, leads to 
better agreement in the forward region, but deviations are observed 
for central jet production.

Uncertainties related to higher-order corrections to the hard process
are most important for exclusive multi-jet final states. 
Similarly, uncertainties related to the resummation procedure 
are most significant in the region where jet production is modelled 
by the parton shower. We expect that both uncertainties can be reduced 
by application of matrix-element parton-shower merging methods at the 
next-to-leading order~\cite{Lavesson:2008ah,Gehrmann:2012yg,Hoeche:2012yf}.
A corresponding analysis is forthcoming.
A different role is played by non-perturbative uncertainties. 
They are most significant in regions where (semi-)soft particle production
dominates over multi-jet effects. They can be reduced only by better
constraints on the non-perturbative dynamics through additional measurements
of underlying event activity and particle flow.

Our analysis should help to better understand the quality of Monte-Carlo
predictions, which are obtained using matching methods like \MCatNLO and \POWHEG.
Ascribing reliable uncertainties to such predictions will remain an
important task in the immediate future of \LHC physics.

\section*{Acknowledgements}

SH's work was supported by the US Department of Energy under contract 
DE--AC02--76SF00515 and in part by the US National Science Foundation, grant 
NSF--PHY--0705682, (The LHC Theory Initiative).  MS's work was supported by 
the Research Executive Agency (REA) of the European Union under the Grant 
Agreement number PITN-GA-2010-264564 (LHCPhenoNet). MS would also like to 
thank the SLAC National Accelerator Laboratory, where parts of this project 
have been completed, for its kind hospitality.

%= bibliography ===================================
\bibliographystyle{bib/amsunsrt_mod}
\bibliography{bib/journal}
%= end ============================================
\end{document}